\documentclass[fleqn,usenatbib]{mnras}

\usepackage{newtxtext,newtxmath}

\usepackage[T1]{fontenc}
\usepackage{hyperref}

\DeclareRobustCommand{\VAN}[3]{#2}
\let\VANthebibliography\thebibliography
\def\thebibliography{\DeclareRobustCommand{\VAN}[3]{##3}\VANthebibliography}

\usepackage{graphicx}	
\usepackage{amsmath}	

\defcitealias{Weaver2024}{W24}
\defcitealias{Claeyssens2023}{C23}

\title[Tracing star formation across cosmic time]{Tracing star formation across cosmic time at tens of parsec-scales in the lensing cluster field Abell 2744.}

\author[A. Claeyssens et al.]{Adélaïde Claeyssens,$^{1}$\thanks{E-mail: adelaide.claeyssens@astro.su.se}
Angela Adamo,$^{1}$
Matteo Messa$^{2}$, Miroslava Dessauges-Zavadsky$^{3}$, Johan Richard$^{4}$
\newauthor
Ivan Kramarenko$^{5}$, Jorryt Matthee$^{5}$, Rohan P. Naidu$^{6}$
\\
$^{1}$The Oskar Klein Centre, Department of Astronomy, Stockholm University, AlbaNova, SE-10691 Stockholm, Sweden\\
$^{2}$INAF -- OAS, Osservatorio di Astrofisica e Scienza dello Spazio di Bologna, via Gobetti 93/3, I-40129 Bologna, Italy\\
$^{3}$Observatoire de Genève, Université de Genève, Versoix, Switzerland\\
$^{4}$Univ Lyon, Univ Lyon1, ENS de Lyon, CNRS, Centre de Recherche Astrophysique de Lyon UMR5574, Saint-Genis-Laval, France\\
$^{5}$Institute of Science and Technology Austria (ISTA), Am Campus 1, 3400 Klosterneuburg, Austria\\
$^{6}$ MIT Kavli Institute for Astrophysics and Space Research, 77 Massachusetts Avenue, Cambridge, 02139, Massachusetts, USA \\
}

\date{Accepted XXX. Received YYY; in original form ZZZ}

\pubyear{2024}

\begin{document}
\label{firstpage}
\pagerange{\pageref{firstpage}--\pageref{lastpage}}
\maketitle

\begin{abstract}
We present a sample of 1956 individual stellar clumps at redshift $0.7<z<10$, detected with JWST/NIRCam in 476 galaxies lensed by the galaxy cluster Abell2744. The lensed clumps present magnifications ranging between $\mu$=1.8 and $\mu$=300. We perform simultaneous size-photometry estimates in 20 JWST/NIRCam median and broad-band filters from 0.7 to 5 $\mu$m. Spectral energy distribution (SED) fitting analyses enable us to recover the physical properties of the clumps. The majority of the clumps are spatially resolved and have effective radii in the range $\rm R_{eff} = 10 - 700 \ pc$. We restrict this first study to the 1751 post-reionisation era clumps with redshift $<5.5$. We find a significant evolution of the average clump ages, SFR, SFR surface densities and metallicity with increasing redshift, while median stellar mass and stellar mass surface densities are similar in the probed redshift range. We observe a strong correlation between the clump properties and the properties of their host galaxies, with more massive galaxies hosting more massive and older clumps. We find that clumps closer to their host galactic center are on average more massive, while their ages do not show clear sign of migration. We find that clumps at cosmic noon sample the upper-mass end of the mass function to higher masses than at $z>3$, reflecting the rapid increase towards the peak of the cosmic star formation history. We conclude that the results achieved over the studied redshift range are in agreement with expectation of in-situ clump formation scenario from large-scale disk fragmentation.

\end{abstract}




\section{Introduction}

Hubble Space Telescope (HST) observations revealed that the majority of the distant galaxies ($z>1$)  present a very irregular morphology dominated by clumpy structures. Initial studies of non-lensed observations, up to redshifts $\sim 3$, have concluded that these clumps are giant star-forming regions with sizes of $\sim1$ kpc (\citealt{Elmegreen2006, Wisnioski2012}). However, HST observations of magnified distant galaxies have shown that when increasing the spatial resolution, kpc-size structures are broken down into increasingly smaller structures. Magnified galaxies exhibit clumps with sizes ranging from 500 to 10 pc, approaching in some cases stellar clusters (\citealt{Johnson2017, DZ2017,Cava2018,Vanzella2019,Mestric2022, Messa2022, Messa2024}). Several statistical studies of distant galaxies with HST (\citealt{Guo2015, Shibuya2016, Sattari2023}) reported that the fraction of galaxies with clumpy structure peaks at cosmic noon ($1<z<3$, at the peak of the cosmic SFR density, \citealt{Madau2014}), suggesting that clump formation could be closely related to the physical conditions under which star formation takes place in galaxies. 

\noindent HST-based studies have focused in understanding how star formation operates at sub-galactic scales by linking clump physical properties to the galaxy mass build-up and evolution.  
Based on their UV-optical colours and spatial distributions within their host galaxies, it has been speculated that  clumps could survive during a several hundreds of Myr, and migrate toward the (proto-)bulge and as such contribute to its formation (\citealt{Elmegreen2009, FS2011,Zanella2019}). Clumps dominate the rest-frame UV morphology and light distribution of galaxies (\citealt{Shibuya2015}).
Redder clumps are also observed and are consistent with old or extincted stellar populations (\citealt{Guo2012, Guo2015}). Studies of local-analogues of highly star-forming galaxies (gas-rich disk and/or characterised by large SFR densities) at $z=1-2$ show that they host clumps with similar sizes and masses to those observed in main sequence galaxies at cosmic noon (\citealt{Fisher2017a, Fisher2017b, Messa2019}).  \\

\noindent Two major clump formation scenarios have been proposed in the literature: in-situ and merger. The first suggests that clumps are formed by gravitational fragmentation due to violent disk instabilities triggered by gas inflows (\citealt{Agertz2009, Dekel2009, Bournaud2010}). The host galactic dynamics are dominated by random motions and turbulent thick disks, suggesting also that clumps are produced by fragmentation in gravitationally unstable disks. The clumpy nature of gas-rich disks has been well-retrieved in many simulations of galaxies in isolated and cosmological context (among many others, see \citealt{Nogushi1999,Agertz2009,Dekel2009,Bournaud2010,Ceverino2012,Hopkins2012, Bournaud2014, Tamburello2015, Behrendt2016, Madelker2017,Fensch2021, Renaud2021b}).   
Simulations of disk galaxies with different gas fractions, \citet{Renaud2021b,Renaud2024} have shown that it is the gas-rich nature of these disks, which allows the formation of local clumpy structures. 
The second scenario suggests that the clumps could be accreted remnants following minor merger events (\citealt{Madelker2017}), and some observational signature supporting this scenario have been reported (\citealt{Zanella2019}). At higher redshift ($z=5.5-9.5$), the First Light simulations have found that most of the clumps are formed by mergers (\citealt{Nakazato2024}). \\

\noindent The survival timescale of clumps is subject to an active debate, with numerical simulations finding either short-lived or long-lived clumps. From one side, short-lived clumps are typically disrupted by stellar feedback in a few tens of Myr (\citealt{Genel2012,Tamburello2015}). Massive star-forming galaxies at $z<2$ in the Illustris-TNG simulations  \citep{Pillepich2019} even rarely or never exhibit giant clumps in the disks, the galaxies are almost always dominated by a regular spiral structure.
On the other hand, several simulations including complex models of stellar feedback (supernovae, radiation pressure, ionization by HII regions) do show the formation of relatively long-lived ($> 100-500$ Myr) giant clumps (\citealt{Perret2014, Bournaud2014, Ceverino2014, Fensch2017, Calura2022,Ceverino2023}). For instance, \citet{Perez2013} show that 100~pc to 1~kpc scale clumps can survive even when strong outflows -- consistent with observations (\citealt{Genzel2011,Newman2012}) -- are released during their star formation activity. \citet{Ceverino2023} showed that, while the majority of this type of clumps are quickly disrupted, within 10-50 Myr, due to strong feedback right after the onset of star formation, the denser clumps can survive longer than the disk dynamical time. \\

\noindent Several numerical simulations have also shown that clumps, formed in the disk, will migrate towards the center of the galaxy within a few hundreds Myr, and ultimately merge, and thus contribute to the formation of a massive bulge (\citealt{Dekel2009, Ceverino2010, Krumholz2010, Cacciato2012, Forbes2012, Forbes2014, Krumholz2018, Dekel2020, Dekel2021, Ceverino2023}). Signs of migration have been reported in the observations of distant clumpy galaxies in some studies. Using the HST CANDELS-GOODS observations of  3187 clumps from 1269 galaxies at $z=0.5-3$ \citep{Guo2018}, \citet{Dekel2021} have measured a significant radial age and mass gradient across the galaxies consistent with their simulated clumps. The mass-distance trend suggests that the clumps could grow in mass during their migration via clumps mergers, or be directly formed more massive in the inner and thicker part of the disk. \\

\noindent Thanks to the unprecedented resolution and sensitivity of the James Webb Space Telescope (JWST), and especially the Near Infrared Camera (NIRCam), reaching AB magnitude of $\sim 30$, more clumps can now be detected within distant galaxies. For instance, \citet{Kalita2024} studied a population of 418 star-forming non-lensed galaxies at $1<z<2$ from the Cosmic Evolution and Epoch of Reionization Survey (CEERS) observations with JWST/NIRCam. They probed clumps mass ranging between $10^7$ and $10^{9.5} \ \rm M_{\odot}$. However, without lensing magnification, this sample is strongly limited in physical spatial resolution, and thus they detect only kpc-scale clumps which are likely composed of multiple blended smaller unresolved structures. \\

\noindent Combining JWST/NIRCam observations with the magnification power of the gravitationally lensing effect allow us to overcome this limitation by detecting large samples of clumpy structures down to $<10$ pc and $10^5 \ \rm M_{\odot}$. \citet{Claeyssens2023} (hereafter \citetalias{Claeyssens2023}) have built the first statistical sample of sub-kpc clumps with JWST/NIRCam (223 clumps from 18 galaxies at $z=1-8.5$) from the early release observations of the lensing cluster SMACS0723. They derived optical rest-frame effective radii from $<10$ to a few hundreds pc and masses ranging from $10^5$ to $10^9$  $\rm M_{\odot}$, overlapping with massive star clusters and stellar complexes in the local universe, but extending also to mass ranges rarely seen in the local universe at sub-hundreds parsec scales. Clump ages range from 1 Myr to 1 Gyr and 45 to 60\% of them are consistent with being gravitationally bound. On average, the dearth of $\sim$Gyr old clumps suggests that survival timescales are typically shorter than that. Moreover, they detected a significant increase in the clump stellar mass surface density with redshift, similarly reported by \cite{Messa2024}. Clumps at high-z can reach stellar densities up to $\rm 10^4-10^5~M_\odot~pc^{-2}$, i.e. higher than typical individual star clusters in the local Universe \citep[e.g.][]{BG2021}.  
Overall, the statistical sample of \citetalias{Claeyssens2023}, along with other JWST studies focusing on individual extremely magnified high-z galaxies \citep[e.g.,][]{vanzella2022_a2744,Vanzella2023,mowla2024,Adamo2024,messa2024_D1T1}, demonstrated that individual star cluster candidates can be detected and studied in these systems, and can be used to trace the recent star formation history of their host galaxies, as well as, under what conditions star formation is taking place. Among the 18 galaxies of \citetalias{Claeyssens2023}, two galaxies -- the Sparkler and the Firework galaxies -- show clump colour distributions and clump locations surrounding their host galaxy that are compatible with being globular cluster-like accreted or formed during merger events. The ages of some of the compact clusters, especially within the Sparkler galaxy at $z=1.38$ are between 1 and 4 Gyr, e.g., consistent with being globular cluster precursors formed around 9-12 Gyr ago (see also \citealt{Mowla2022,Adamo2023} for detailed studies of the Sparkler galaxy). 

\noindent These first studies have shown the great potential of JWST observations for detecting and characterising both hundreds pc scale stellar regions and individual star clusters, and thus open a new window on the conditions under which star formation occurs and star clusters form in rapidly evolving galaxies. High spatial resolution clumpy galaxies studies with JWST and HST have so far focused on small samples (a few hundreds) or single-galaxy analyses between $z=1$ and $z=10$. To pursue our investigations on the nature of these clumps, their formation and evolution processes, and their impact on the galaxy evolution, larger samples are necessary.\\

\noindent We present in this paper a new sample of 474 unique galaxies at $z=0.7-10$ magnified by the lensing cluster A2744. Taking advantage of the numerous datasets of multiple observing JWST/NIRCam programs executed on this galaxy cluster region (20 bands in total among medium and broad band filters), we are able to get robust properties, such as age, SFR, mass, and size for 1956 individual clumps. In Sect.~\ref{sec:data} we present the different observations exploited. Sect.~\ref{sec:method} describes the galaxy selection, and the measurements of the galaxy and clump properties. We report on the main physical properties in Sect.~\ref{sec:results}, and discuss different clump formation and evolution scenarios in Sect.~\ref{sec:discussion}.

\section{Data}
\label{sec:data}

\subsection{HST observations}
\label{sec:HST_obs}

The core of the galaxy cluster Abell 2744 (hereafter A2744) has been covered by deep HST imaging observations executed between 2013 Oct 25 and 2014 Jul 1 as part of the Frontier Fields HST program (ID: 13495, P.I: J. Lotz, \citealt{Lotz2017}). The HST observations include imaging in three filters (F435W, F606W, F814W) of the Advanced Camera for Surveys (ACS), and in four filters (F105W, F125W, F140W, F160W) of the Wide-Field Camera Three (WCF3). In total 280 orbits of data have been acquired reaching a 5-$\sigma$ limiting magnitude AB$\sim$29 in each filter. A wider area has been covered between 2018 and 2020 by shallower observations from the BUFFALO program \citep{Steinhardt2020}.

\subsection {JWST observations}
\label{sec:JWST_obs}

\subsubsection {JWST/NIRCam}

The A2744 core region has been imaged with the Near-Infrared Camera (NIRCam) on the JWST by two different programs. The UNCOVER program (GO 2561, P.Is. Labbé and Bezanson, and presented in \citealt{Bezanson2022}) consists of short wavelength (SW) imaging in F115W, F150W, F200W, combined with long wavelength (LW) imaging in 3 other broadband F277W, F356W, F444W, and one medium band filter (F410M)\footnote{publicly available on the UNCOVER webpage (\href{test}{https://jwst-uncover.github.io})}. 
 The core of A2744 has also been observed during Cycle 2 under the program named Mega-Science (GO 4111, P.I. Suess). Mega-Science observations \citet{Suess2024}, imaged the cluster region in 10 medium-band filters (F140M, F162M, F182M, F210M, F250M, F300M, F335M, F360M, F430M, F460M, F480M) and two broad-band filters (F070W, F090W). The data we are using in this study combine these two datasets, covering a total field of view of $\sim 40 \ \rm arcmin^2$. The data were reduced using the grizli software \footnote{https://github.com/gbrammer/grizli} (Brammer et al. in prep.) and were drizzled into mosaics with a common pixel scale of $0.04''$ for the HST and JWST LW filters and $0.02''$ for the JWST SW filters with exposure times ranging between 2 to 5 hours per filter.
In total, the JWST/NIRCam observations of A2744 are available in 20 different broad and medium-band filters from 0.7 to 4.8 $\mu$m, making this cluster the best sampled with JWST/NIRCam photometric data so far (cf Tab.~\ref{tab:table_observations}).

\subsection {VLT/MUSE data}
\label{sec:MUSE_data}

A2744 has also been observed with the Multi Unit Spectrographic Explorer (MUSE) at the Very Large Telescope (VLT) between September 2014 and October 2015 as part of the GTO Program 094.A-0115 (PI: J. Richard) and during the program 109.24EZ.001 (PI: C. Mason). The data consist of a $2\times2$ mosaic of MUSE pointings, which has been designed to cover the entire multiple image area of the core of the cluster, as well as 3 shallower pointings in the outskirts. The depth of the observations varies from 0.75 to 7 hours per pointing across the field. The data reduction of these MUSE data follows the details presented in  \citet{Richard2021}, and the catalogue of spectroscopic redshifts  based on the combination of HST and MUSE data in the central region of the cluster is detailed in \citet{Mahler2018} and \citet{Richard2021}. In short, we extract and inspect spectra based on HST identifications as well as the Muselet tool (MUSE Line Emission Tracker, \citealt{Piqueras2019}), which makes use of SExtractor \citep{BertinArnouts1996} to detect emission line objects in the MUSE Narrow Band continuum subtracted images. Based on the emission and absorption lines detected in each source, a spectroscopic redshift has been estimated for 506 objects \citep{Richard2021}, including 471 at high confidence\footnote{Full spectroscopic catalog available at \url{https://cral-perso.univ-lyon1.fr/labo/perso/ johan.richard/MUSE_data_release/}} (with a high redshift confidence).

\section{Method}
\label{sec:method}

\subsection{Galaxy selection}
\label{sec:galaxy_selection}

\begin{figure*}
	\includegraphics[width=18cm]{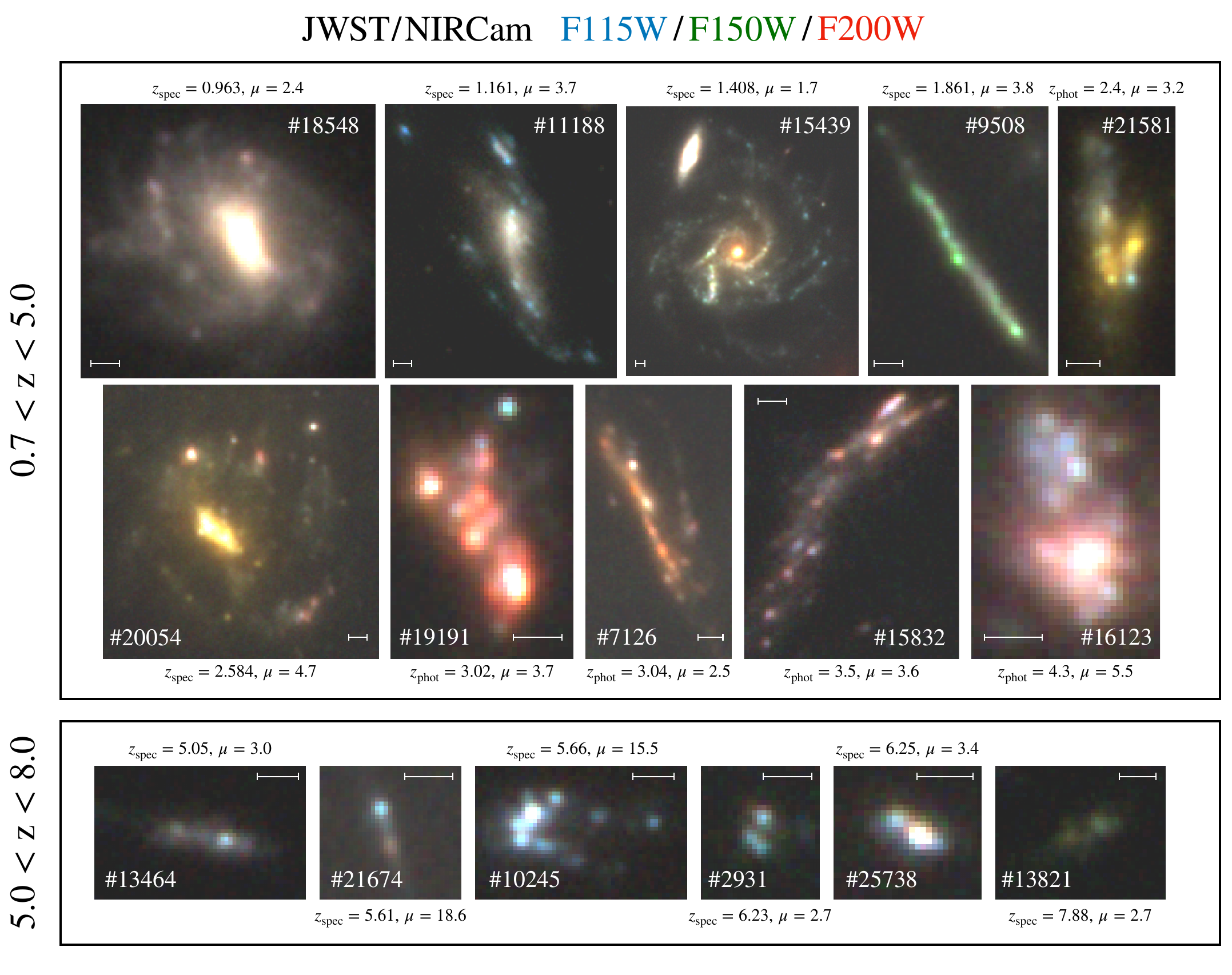}
    \caption{RGB JWST/NIRCam images (with R=F200W, G=F150W and B=F115W) of 16 galaxies representative of the sample diversity. The ID of each galaxy (from the UNCOVER catalog) is indicated in white on each image. The redshift and average magnification are indicated in black below or above each image. The white scale indicates 0.2 arcsecond.}
    \label{fig:rgb_gal}
\end{figure*}

We selected galaxies from the photometric catalogue of \citet{Weaver2024} (hereafter W24, data release 3 of the UNCOVER program\footnote{https://jwst-uncover.github.io/DR3.html}) based on a combination of the archival 7 HST bands with the 20 JWST imaging from the different programs presented in Sect.~\ref{sec:JWST_obs}. Galaxy candidates have been detected on a ultra-deep noise-equalized stacked detection image combining three LW filters: F277W, F356W and F444W. In total, they have detected 74,020 systems including foreground galaxies, galaxy cluster members, and background lensed galaxies. The public data release contains HST and JWST photometry of all the extracted objects and redshift estimates. Aperture photometry has been measured on PSF-matched images to the JWST/F444W resolution after they have been cleaned from contaminating light from bright cluster galaxies (BCG) and intra-cluster light (ICL). Photometric redshifts have been computed using EAzY \citep{Brammer2008}.
The UNCOVER photometric catalogue is among the deepest catalogues publicly available (reaching effective 5$\sigma$ (observed) depths greater than 29 AB in all 15 bands). When combining this survey depth with magnification produced by the strong lensing, this catalogue represents our deepest view into the Universe to date. 

\subsubsection{Selection criteria from W24}
\label{sec:selection_W24}

In order to build a sample of distant clumpy galaxies (i.e., with at least 2 compact and brights structures detected within the galaxy) with secure redshift estimates, we initially selected from the complete \citetalias{Weaver2024} catalogue the sources with: 
\begin{itemize}
    \item A spectroscopic redshift estimation reported in the catalogue from \citetalias{Weaver2024} (obtained from MUSE redshift catalogues presented in \citealt{Richard2021}) or a photometric redshift estimation from \citetalias{Weaver2024} with a 68\% uncertainty lower than $0.25$ to get only robust redshift estimations. These photometric redshifts are in very good agreement with spectroscopic redshifts estimations, when available, see Sect. \ref{sec:spec_z} and \citealt{ALT})
    \item A redshift estimation with $z>0.7$ to get distant galaxies and avoid contamination from the galaxy cluster members.
    \item A reported magnification factor $\mu_{\rm gal}>2$ in \citetalias{Weaver2024} to get an minimal factor of 2 boost in physical resolution and luminosity. The magnification estimates presented in \citetalias{Weaver2024} are computed from the analytic strong lensing mass model presented in \citet{Furtak2023}. This lens model is constrained by 141 multiple images from 48 individual sources and reach an observed  $\rm RMS$ of 0.51 arcseconds.
\end{itemize}  
A total of 851 candidates satisfied these automatic selection criteria. They were subsequently visually inspected by three team members to remove interlopers: saturated stars, galaxy cluster members or intracluster globular clusters, or multiple detections of extended objects in the segmentation map. During visual inspection we also excluded galaxy candidates that suffered from strong contamination due to proximity to bright cluster members, or by foreground objects which strongly affect the quality of the photometry. For post-reionisation era galaxies (at $z<5.5$), we kept only galaxies with at least two clumps. At $z>5.5$ we selected also the galaxies with only one clump identified. Our total final sample consists of 484 galaxies (474 unique galaxies after removing the multiple images of the same objects) with $0.7<z<10$.

\noindent Once the final galaxy catalogue has been established, we verified the photometric redshift estimates produced by \citetalias{Weaver2024} against the publicly available catalogue on the DAWN JWST Archive (DJA\footnote{https://dawn-cph.github.io/dja/blog/2024/03/01/nirspec-extractions-v2/}) produced using all the available JWST/NIRCam bands (medium and broad bands, cf Sect.~\ref{sec:JWST_obs}) therefore providing very robust redshift estimates (see \citealt{Suess2024}). We find a very good agreement between the two catalogues.

\subsubsection{Additional spectroscopic redshifts}
\label{sec:spec_z}

\noindent Among the 484 selected images (from 474 unique galaxies, 16 are presented in the Fig.~\ref{fig:rgb_gal}), 266 have a spectroscopic redshift estimate (i.e. $\sim 55 \%$) accessible in the literature. These spectroscopic redshifts come in majority  from MUSE observations (124 galaxies, \citealt{Richard2021}, see Sect.~\ref{sec:MUSE_data}). Aside, eight redshifts ($z>6$) have been measured by \citet{Atek2023} with NIRSpec MSA follow-up observations. Seven redshifts ($z\sim 7.88$) come from NIRSpec MSA observations presented in \citet{Morishita2023}. One galaxy redshift come from NIRSpec observations presented in \citet{RB2023}. Six redshifts have been measured from the GLASS NIRISS observations \citep{He2024}. Finally, 120 redshifts have been estimated from the NIRCam/GRISM observations (program  3516, "All the Little Things", PI: Matthee and Naidu, \citealt{ALT}). 

\noindent Given the multiple selection criteria used, our selected galaxy sample is not complete to a specific faintness level, but contains galaxies with secure redshift estimates and a minimum magnification of a factor 2. That enables us to study clump physical properties at at least 1.4 times the physical resolution at a given redshift and detect at least twice fainter clumps that would have been missed in normal field studies.

\subsection{Galaxy SED fitting}
\label{sec:galaxy_SED}

In order to access the global host galaxy physical properties when analysing the clumps, we analysed the galaxy SED photometry with synthetic stellar population models. This was a necessary step as publicly available catalogues do not provide SED fit output information. Moreover, in some cases, photometric entries of multiple segments from the same galaxy are available, which required their combination to obtain the total flux in each bands. We also decided to use the extracted photometry from only one catalogue \citepalias{Weaver2024} to ensure that the photometry, error analysis, and local background correction are all performed in a homogeneous self-consistent way. 

\noindent We used the \citetalias{Weaver2024} HST and JWST photometry (27 filters in total) to perform SED fitting of our 474 galaxies using the SED fitting code BAGPIPES (Bayesian Analysis of Galaxies for Physical Inference and Parameter EStimation by \citealt{Carnall2018, Carnall2020}). We fixed the redshift of each galaxy to the best estimate available in our catalogue (see Sect.~\ref{sec:galaxy_selection}). We fitted the galaxy SED using three different star formation histories (SFH) from BAGPIPES. We first used a constant SFH (hereafter BP\_constant). For this SFH we let 8 parameters free: the stellar mass formed ($\rm M_{\rm formed}$ with a uniform prior between $10^3$ and $10^{15} \ \rm M_{\odot}$), the starting time of the star formation ($t_{\rm start}$) with a uniform prior between the redshift of the source and the age of the Universe, the metallicity ($Z$) with a logarithmic prior between $10^{-4}$ and 4 $\rm Z_{\odot}$, the dust content ($\rm A_V$), and the ionisation parameter ($\rm log(U)$) with a prior between $-4$ and $-2$. For every SFH, we assumed the \citet{Calzetti2000} dust attenuation law with $A_V$ free with a uniform prior between 0 and 4 magnitudes (ABmag).

\noindent We also tested an exponential declining SFH (hereafter BP\_exp) where the SFR is parameterised as: 
\begin{equation*}
\rm SFR(t) \propto 
\begin{cases}
e^{-\frac{t-T_0}{\tau}} & \quad t>T_0 \\
0 & \quad t<T_0
\end{cases}
\end{equation*}
In addition to the free parameters of the BP\_constant SFH (for which we adopted the same priors), this SFH includes the timescale of the decline, $\tau$, for which we adopt a uniform linear prior between 1 Myr and 10 Gyr; the start of star formation, described in BP\_constant by $t_{\rm start}$ is now parametrised by $T_0$.
Finally, we also used a double-power law form (hereafter BP\_dblplaw) for which: 
\begin{equation*}
    \rm SFR(t) \propto \biggl[ \biggl(\frac{t}{\tau}\biggr)^\alpha + \biggl(\frac{t}{\tau}\biggr)^{-\beta} \biggr]^{-1},
\end{equation*}
where $\alpha$ and $\beta$ are the falling and rising slopes, respectively, and $\tau$ is related to the time at which star formation peaks. The new free parameters of this SFH are $\alpha$ and $\beta$, for which we gave very broad logarithmic priors between $-3$ and 7, respectively.

\noindent For the three SFHs, the SFR was estimated from the posterior distributions of parameters over the last 100 Myr. 

\noindent The returned SED fit of each galaxy was visually inspected. For 33 out of the 474 galaxies, we analysed the galaxy SED excluding the HST photometry due to the lack of robust detections of these galaxies in any of the HST bands.

\noindent As reference, in our analysis we employ the best-fit values of each parameter $x_{\rm best}$. The uncertainties associated to the best values are the 16\% and 84\% values ($x_{16}$ and $x_{84}$) extracted from the sampled posterior distributions. The resulting uncertainties are thus asymmetric and defined as $\Delta_{\rm left}=x_{\rm best}-x_{16}$ and $\Delta_{\rm right}=x_{84}-x_{\rm best}$.

\begin{figure}
	\includegraphics[width=8cm]{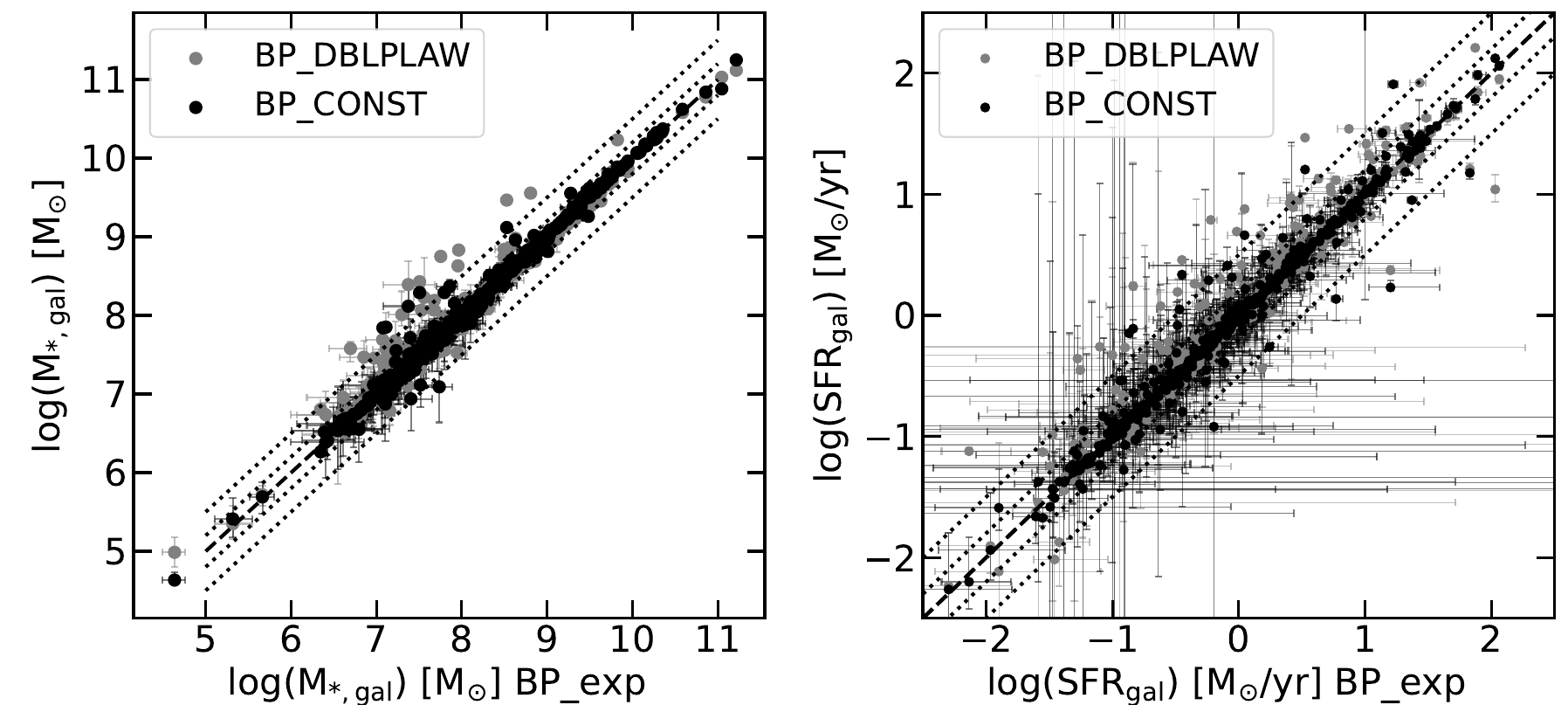}
    \caption{Left: Stellar mass of the galaxies obtained from bagpipes SED fitting with BP\_DBLPLAW and BP\_CONST SFHs as a function of the stellar mass obtained with BP\_exp. Right: SFR of the galaxies obtained from bagpipes SED fitting with BP\_DBLPLAW and BP\_CONST SFHs as a function of the SFR obtained with BP\_exp. The dashed line represent the 1:1 relation. The dotted lines represent the deviation from -0.5; -0.2, +0.2 and +0.5 dex.}
    \label{fig:Sample_galaxies_SFH}
\end{figure}

\noindent Fig.~\ref{fig:Sample_galaxies_SFH} shows the stellar masses ($\rm M_{*, gal}$) and star formation rates ($\rm SFR_{gal}$) estimated with BP\_dblplaw and BP\_constant as a function of the values obtained from BP\_exp. We find a very good agreement between the three SFHs both in stellar mass and SFR. As showed by \citet{Pacifici2023}, the stellar mass is the most robust parameter. For the rest of our analysis we will use the BP\_exp SFH for the galaxies as a reference model.

\begin{figure}
	\includegraphics[width=8.5cm]{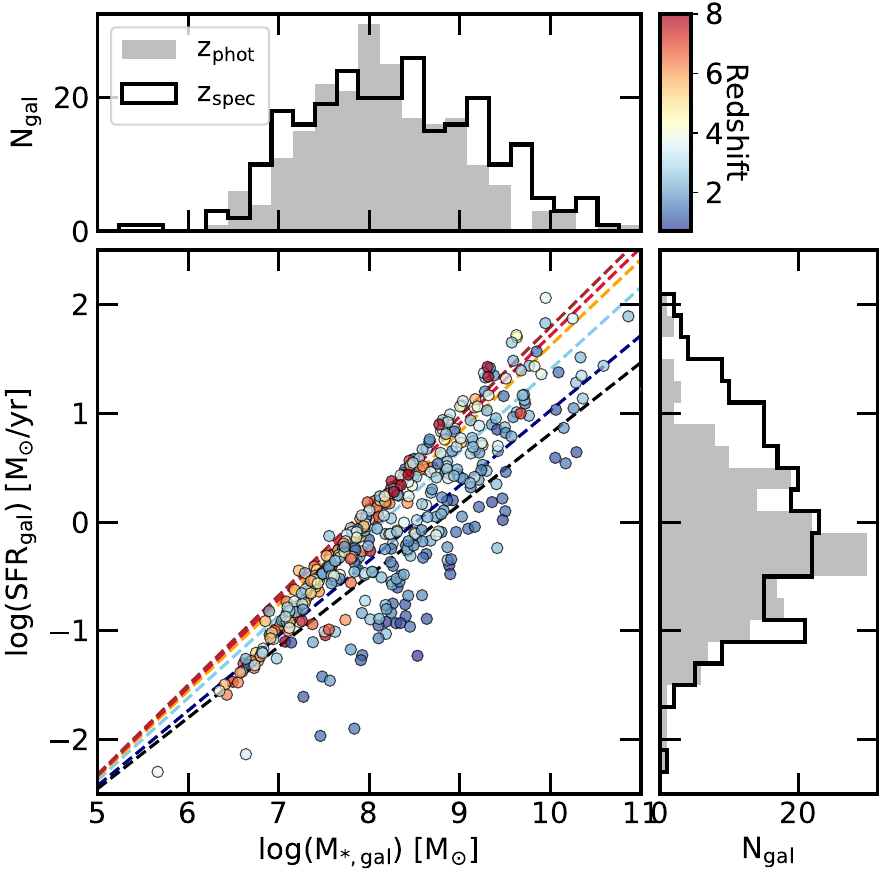}
    \caption{SFR of the galaxies as a function of the stellar mass, obtained from the BAGPIPES SED fitting with an exponential declining SFH (BP\_exp). The points are colour-coded with redshift. The top and left histograms show the stellar mass and SFR distributions of the galaxies with a photometric redshift (in grey) and spectroscopic redshift (in black). The dashed lines represent the star-forming galaxy main sequence (\citealt{Speagle2014}) at $z=0.7; 1; 2; 3.5; 5; 8$. }
    \label{fig:Sample_galaxies_MS}
\end{figure}

\noindent Fig.~\ref{fig:Sample_galaxies_MS} presents $\rm SFR_{gal}$ versus $\rm M_{*, gal}$ obtained from the BP\_exp SFH, colour-coded by the redshift of the galaxies. The sample covers a wide range of galactic stellar masses from $10^6$ to $10^{11} \ \rm M_{\odot}$. We observe that the most massive galaxies ($\rm M_{*,gal}>10^{9} \ \rm M_{\odot}$) are mainly at $z<3$, while at redshift $z>6$, we only detect galaxies $\rm M_{*,gal}<10^{9} \ \rm M_{\odot}$. The galaxies cover a wide range of SFR values from $\rm 10^{-2}$ to $\rm 10^2 \ M_{*}/yr$ at all redshifts. Overall, the galaxy sample distributes around the star-forming galaxy main sequence as a function of redshift as measured in \citet{Speagle2014}.

\begin{figure}
	\includegraphics[width=9cm]{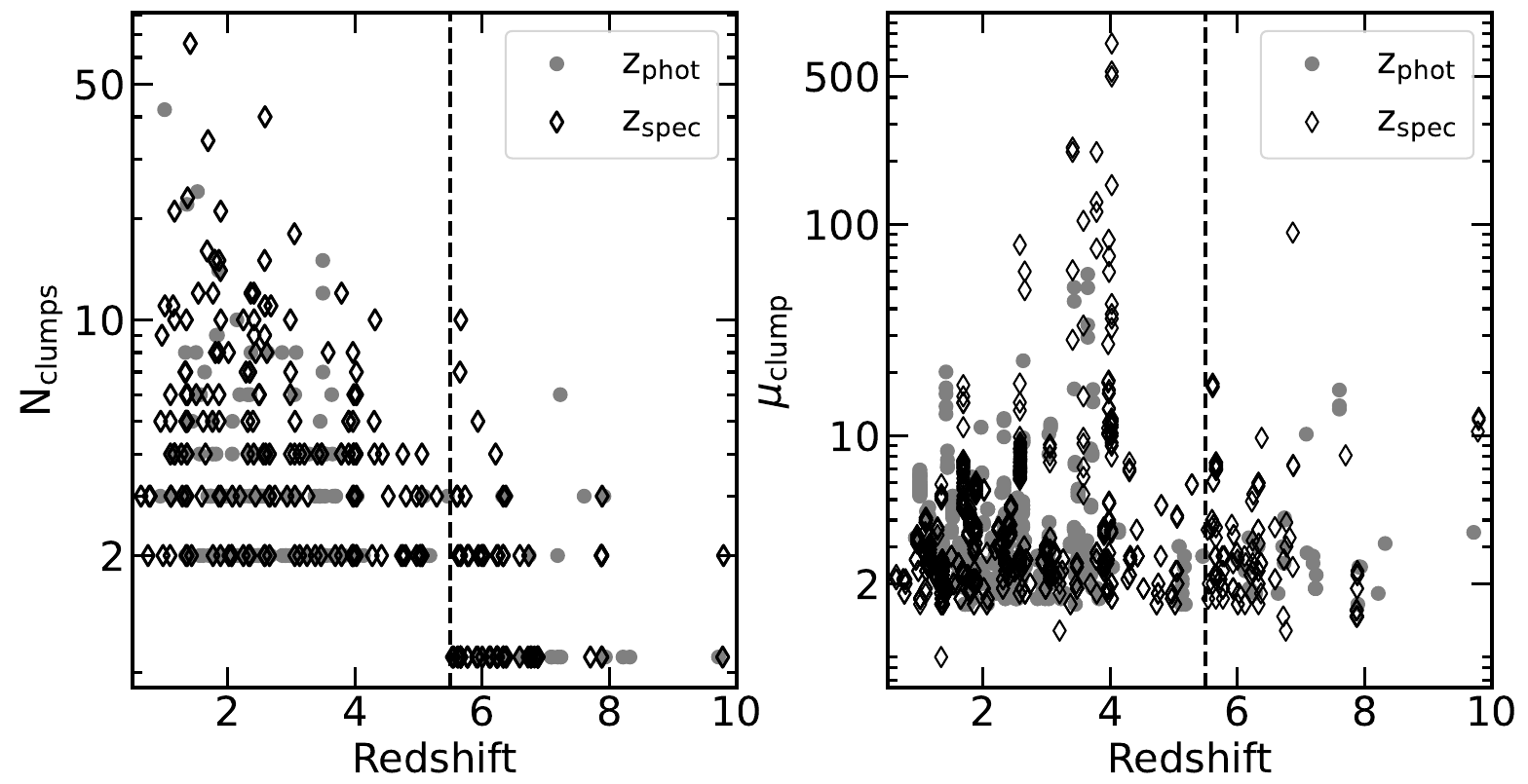}
    \caption{Left: Number of clumps detected per galaxy as a function of redshift. Right: Clump magnification as a function of redshift. In the two panels, the grey points represent the galaxies with photometric redshifts and the black diamonds the galaxies with spectroscopic redshifts.}
    \label{fig:Sample_galaxies_clumps}
\end{figure}

\subsection{Clumps detection}
\label{sec:clumps_detection}

To improve the detection of clumps, we created cutouts around the host galaxies, using RGB frames based on SW and LW NIRCam filter combinations to maximise the detection of the diffuse light of the galaxy. 
In crowed regions where multiple objects are detected very closely to each other, we used the segmentation map and full redshift catalogue provided by \citetalias{Weaver2024} (on which our initial photometric catalogue has been built, see Sect.~\ref{sec:galaxy_selection}) to disentangle the structures which belong to each galaxy. When a galaxy is detected in the MUSE data, we used both MUSE narrow-band images of the detected emission lines and JWST/NIRCam color images to determine its extent. 

\noindent We then performed an automatic clump detection using Sextractor (through the python package {\tt sep}, \citealt{Barbary2016}) on the filter selected to be the reference.  The reference filter is varying according to the redshift to probe the optical rest-frame wavelengths and use the most sensitive and higher resolution filters. We use F150W for $z<2.6$, F200W for $2.6<z<6$, and F277W for $6>z$. After numerous trial the following parameters produced the most complete clump extraction:  pixel threshold of $2 \times \rm Var$ with Var the variance, minimum number of contiguous pixels equal to 4, a minimum contrast ratio used for object deblending of 0.005, and a number of thresholds used for object deblending of 64.  We performed a visual inspection of each galaxy and manually modified the clump detection when needed by adding non-detected clumps and/or removing double detections of the same clumps. With this inspection, we ensured that all the visible compact and bright structures are detected, that there are no multiple detections of the same sources, and that the choice of the reference filter for the clump detection is adapted to each galaxy. The final clump catalogue contains in total 1956 clumps extracted from the 476 individual galaxies (in total 2004 clumps extracted from 484 galaxy images). The selected galaxies host between 2 and 65 clumps (cf Fig. \ref{fig:Sample_galaxies_clumps}) at $z<5.5$. We included 1-clump galaxies at $z>5.5$ as multiple-clumps galaxies are less numerous. These clumps will be studies in an future analysis.

\subsection{Magnitude limits}
\label{sec:magnitude_limits}

For each JWST/NIRCam filter we determine the intrinsic magnitude limit ($\rm mag_{lim}$, reported in Table~1) corresponding to the median flux of a PSF aperture measured in 3,000 empty regions of the sky in each filter far from the cores of the cluster. This minimum flux is converted to an AB magnitude and used as an upper limit when performing clump SED analyses.

\begin{table} 
 
\begin{tabular}{llllll} 
Filter &  $\rm {mag}_{\rm {lim}}$  & $\rm E(B-V)_{\rm MW}$ &  PSF FWHM & Max exposure \\ 
 & AB mag  & & [arcsec] & [seconds]\\ 

\hline 
\hline 

F070W &  29.0 & 0.025  & 0.052 & 10,300 \\ 
F090W &  29.3 & 0.016  & 0.056 & 54,428 \\ 
F115W &  29.7 & 0.011  & 0.054 & 55,395\\ 
F140M &  28.8 & 0.008  & 0.060 & 10,300 \\ 
F150W &  29.8 & 0.007  & 0.058 & 33,115\\ 
F162M &  28.7 & 0.006  & 0.064 & 10,300 \\
F182M &  29.3 & 0.005  & 0.066 & 19,574 \\ 
F200W &  29.8 & 0.005  & 0.064 & 23,857 \\ 
F210M &  29.2 & 0.004  & 0.072 & 27,636 \\ 
F250M &  28.0 & 0.003  & 0.112 & 9,270 \\
F277W &  29.6 & 0.003  & 0.112 & 27,967 \\
F300M &  28.4 & 0.003  & 0.120 & 11,583 \\
F335M &  28.5 & 0.002  & 0.128 & 11,840 \\
F356W &  29.3 & 0.002  & 0.124 & 28,599 \\
F360M &  28.4 & 0.002  & 0.136 & 12,360 \\
F410M &  28.6 & 0.002  & 0.144 & 17,893 \\
F430M &  27.9 & 0.002  & 0.150 & 9,270 \\
F444W &  29.4 & 0.002  & 0.160 & 82,482 \\
F460M &  27.5 & 0.002  & 0.160 & 12,354 \\
F480M &  27.7 & 0.002  & 0.160 & 11,845 \\
\hline
\end{tabular} 
\caption{Calibration details of JWST/NIRCam images. The magnitude limits reported in the second column have been measured at 3$\sigma$ in PSF-like apertures in each filter. Due to the observation strategies of the different programs, the exposure time varies across the field. The magnitude limits presented here are averaged over the area where galaxies have been selected, therefore they do not reflect the maximum exposure time. The third column gives the galactic extinction we have corrected from the clump photometry in each filter. The fourth column gives the PSF FWHM (measured on a stack of at least 5 isolated and unsaturated stars). The last column gives the maximum exposure time reached  in the central area of the cluster where galaxies have been selected.}
\label{tab:table_observations} 
\end{table}

\subsection{Clumps photometry}
\label{sec:clumps_photometry}

Since stellar clumps will appear with different sizes and resolutions in the different galaxies depending on redshift and magnification factors, fixed aperture photometry would produce significant contamination and might introduce strong systematics to the photometry. Therefore, we simultaneously fit the clump sizes and fluxes using the method developed by \citet{Messa2019} and later improved and adapted by \citet{Messa2022} to take into account the lens models. This method has been widely used and validated on recent JWST/NIRCam studies of clumps in lensed galaxies (\citealt{Claeyssens2023, Vanzella2023, Adamo2024, messa2024_D1T1}).

\noindent We determine clump sizes and fluxes in two filters depending on redshift. We use the JWST/NIRCam F150W and F200W filters for galaxies with $z<6$ and F200W and F277W filters for galaxies with $z>6$. These filters are the most sensitive with a high spatial resolution. We assume that clumps can be modelled with an elliptical 2D Gaussian profile convolved with the instrumental PSF in the image plane, and a spatially-varying local background emission which includes the diffuse light from the galaxy and that we want to remove to get only the clump flux. 
The 2D Gaussian profile is parameterised by the clump centre (x0 and y0), the minor axis standard deviation ($x_{rm std}$), the axis ratio ($\epsilon$), the positional angle ($\alpha$, describing the orientation of the ellipses), and the total flux ($f$). The background is parameterised by a $1^{st}$ degree polynomial function, described by three free parameters.

\noindent We model the PSF of each filter included in the analysis from a stack of 5 bright, isolated and non-saturated stars, detected within the field of view of the galaxy cluster region. The HST or JWST image cutouts of each star are combined to create a $1'' \times 1''$ averaged image of the PSF. Following the method by \citet{Messa2022}, we determine an analytical expression of the PSF shape in each band, by fitting the resulting PSF image with an analytical function described by the combination of a Moffat, Gaussian, and 4th-degree power-law profiles. This fit provides a good description of the PSF up to a radius of $\sim0.5''$, which is significantly larger than the physical region used to fit the clump light distribution. This size also corresponds to the physical scale to which aperture correction is estimated. The PSF FWHM of each filter are presented in Table~\ref{tab:table_observations}. \\

\noindent The spatial fit is performed on a cutout image (9$\times$9 pixels) centred on each clump, using a least-squared approach via the python package {\tt lmfit} by \citet{Newville2021}. We tested different cutout sizes, with the $\rm 9\times9$ box size performing best in most of the cases by producing the least residuals.  We point out that a  9$\times$9 cutout encloses the peak of the clumps and at least 95\% of the flux within the clumps and enough background to estimate the underlying flux. Indeed, as presented in the Table\ref{tab:table_observations}, the stellar PSF FWHM are ranging from 2.5 to 4 pixels max. When two clumps are separated by less than 4 pixels in the F150W images (i.e. $0.08''$), we fitted the two clumps simultaneously using a 13$\times$13 pixels cutout image. This simultaneous fit on a larger box size produced better results than fitting the two nearby clumps individually. During testing, the clump sizes determined by independently fitting a second nearby filter frame produced similar results (as shown in \citetalias{Claeyssens2023} using the same method). Therefore, we choose as a reference for the clumps shape, the resulting 2D Gaussian parameters produced in the detection reference filter selected for each galaxy (corresponding to the filter with the best detection of all the clumps, see Sect.~\ref{sec:clumps_detection}). 
Once the Gaussian shape (defined by $x_{\rm std}$, $\epsilon$, and $\alpha$ parameters) is fixed in the reference filter fit, we convolve the best model to the stellar PSF of the other band where we fitted the clump cutouts on the other bands (listed in the Table.~\ref{tab:table_observations}) by treating the flux $f$ and the local background as free parameters. We also allow the position of the centre (x0, y0) to vary by 1 pixel maximum, to account for drizzling and shifts due to different pixel scales among the data. This assumption ensures that the clumps have intrinsically the same sizes and morphology in all the filters and reduces the risk of including flux originating in areas surrounding the clumps.

\begin{figure*}
	\includegraphics[width=18cm]{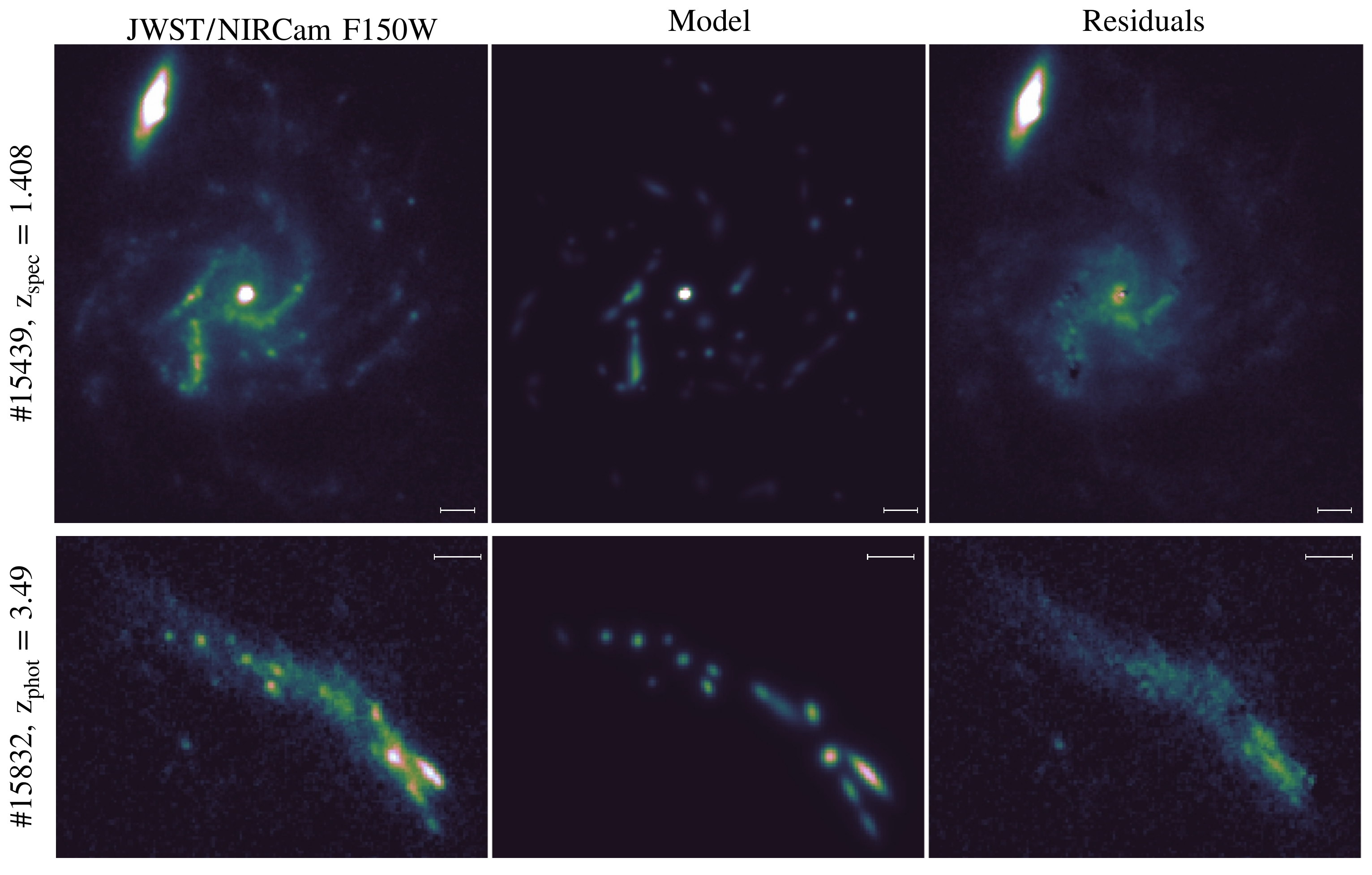}
    \caption{Best-fit photometry for two galaxies. From left to right: Observed image in the reference filter (F150W), best-fit clump model and residual images of the two galaxies at $z=1.408$ (top row) and $z=3.49$ (bottom row). The white scales indicate $0.2''$.}
    \label{fig:best_fit_phot}
\end{figure*}

\noindent Fig.~\ref{fig:best_fit_phot} shows the best-fit photometry for two galaxies (\#15439 at $z=1.408$ with 66 clumps and \#15832 at $z=3.49$ with 15 clumps). The method we use allows to measure the flux contained in each clump while preserving the diffuse light from the galaxy, as shown in the residual image frames.

\subsection{Clumps SED fitting}
\label{sec:clumps_SED}

\begin{figure*}
	\includegraphics[width=18cm]{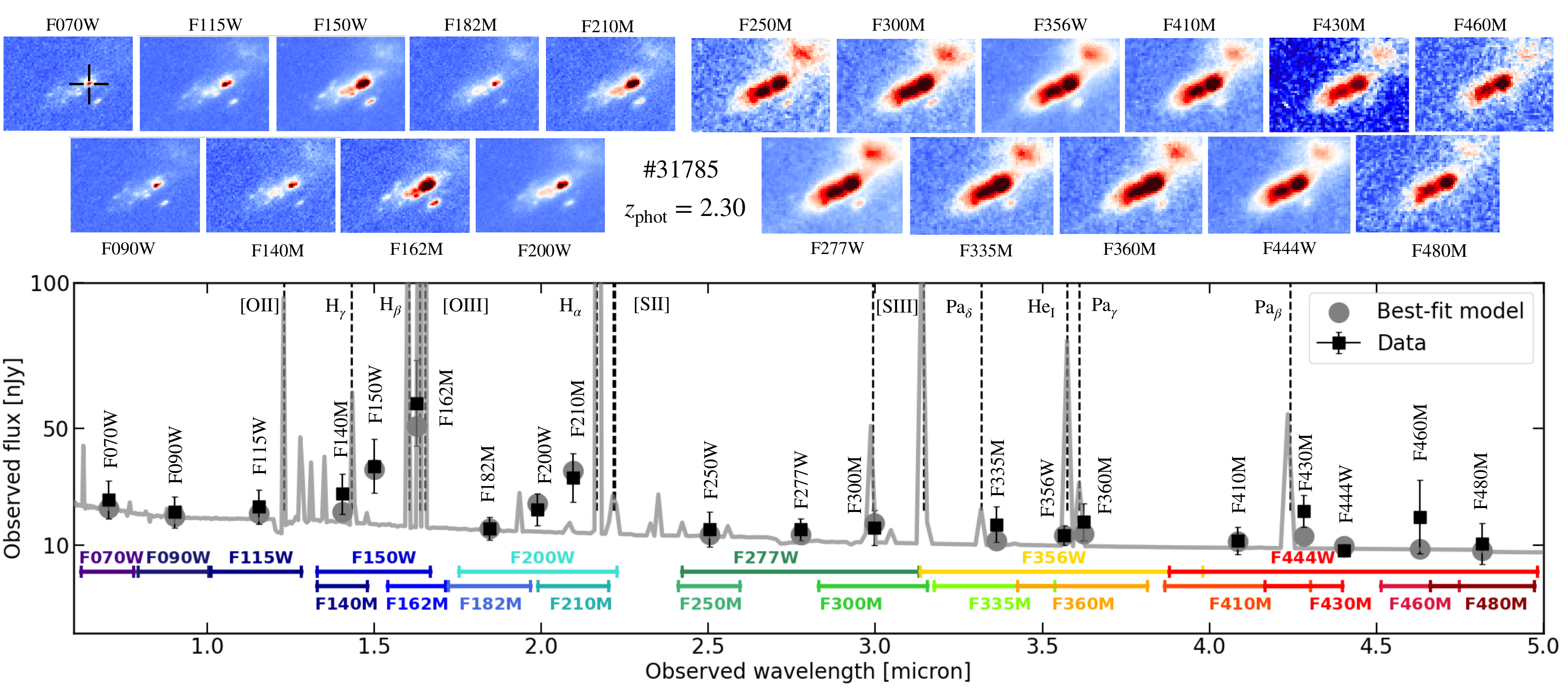}
    \caption{Best BAGPIPES SED fit for the BP\_exp SFH model for the central clump of the galaxy \#31785 at $\rm z=2.30$. We show the image of the galaxy in every filter with the same flux scale, as well as the full best-fit spectrum (in grey) and the observed photometry (in black). The dashed lines indicate the different strong emission lines constrained thanks to their detection in the multiple medium-band filters and their combination with broad-band photometry. We indicate in blue and red the half power wavelengths of each NIRCam filter at the bottom of the spectrum. }
    \label{fig:fig_filters}
\end{figure*}

The final photometric catalogue is then used as input to perform the SED analysis of the clumps and estimate their ages, masses, metallicity, SFR, and extinction. We use BAGPIPES to perform the clump SED fitting analysis. 
Given the small sizes of the clumps ($< 700 \ \rm pc$), we assume relatively short SFHs for these structures, using exponential declining SFH (BP\_exp) with different values of $\tau$. We tested $4$ different SFHs with $\tau = 1 \ \rm Myr$ (hereafter BP\_exp\_1Myr), $\tau = 10 \ \rm Myr$ (BP\_exp\_10Myr), and $\tau = 100 \ \rm Myr$ (BP\_exp\_100Myr), and one with $\tau$ free (BP\_exp). When we compare the outputs of these 4 models, we notice that all of them produce good fits with similar reduced $\chi^2$, meaning that we can not disentangle the most suitable SFH with the available data. In general, short bursts perform equally well when reproducing the clumps light. For all the SFHs the free parameters are the age ($T_0$) with a uniform prior between the redshift of the source and the age of the Universe, the mass formed ($\rm M_{\rm formed}$) with a uniform prior between $10^3$ and $10^{15} \ \rm M_{\odot}$, the metallicity ($Z$) with a logarithmic prior between $10^{-4}$ and 3 $\rm Z_{\odot}$, the dust content ($\rm A_V$) with a uniform linear prior between 0 and 4 (assuming the Calzetti dust law), and the ionisation parameter ($\rm log(U)$) with a prior between $-4$ and $-2$ (and $\tau$ with a uniform linear prior between 1 Myr and 10 Gyr for the last model).  The SFR of the clumps is estimated over the last 10 Myr of the SFH. In total we fit between 5 and 6 free parameters depending on the SFHs, with 20 data-points (i.e. 20 JWST/NIRCam filters). 
In particular, thanks to the medium-band filters, we can capture the strength of emission lines, depending on redshift, for several galaxies (see an example in Fig.~\ref{fig:fig_filters}), which assures a robust constraint on these 5 parameters. We use the magnitude limits reported in Table~\ref{tab:table_observations} as upper limits on the photometry for the SED fit process. When a measured clump magnitude was lower than the corresponding magnitude limit in one filter, we put the flux at $0$ with an uncertainty reaching $2\times \rm f_{lim}$, with $\rm f_{lim}$ being the flux corresponding to the magnitude limit.  We define the uncertainties with the same method as for the galaxy SED fitting (see Sect.~\ref{sec:galaxy_SED}).

\begin{figure*}
	\includegraphics[width=18cm]{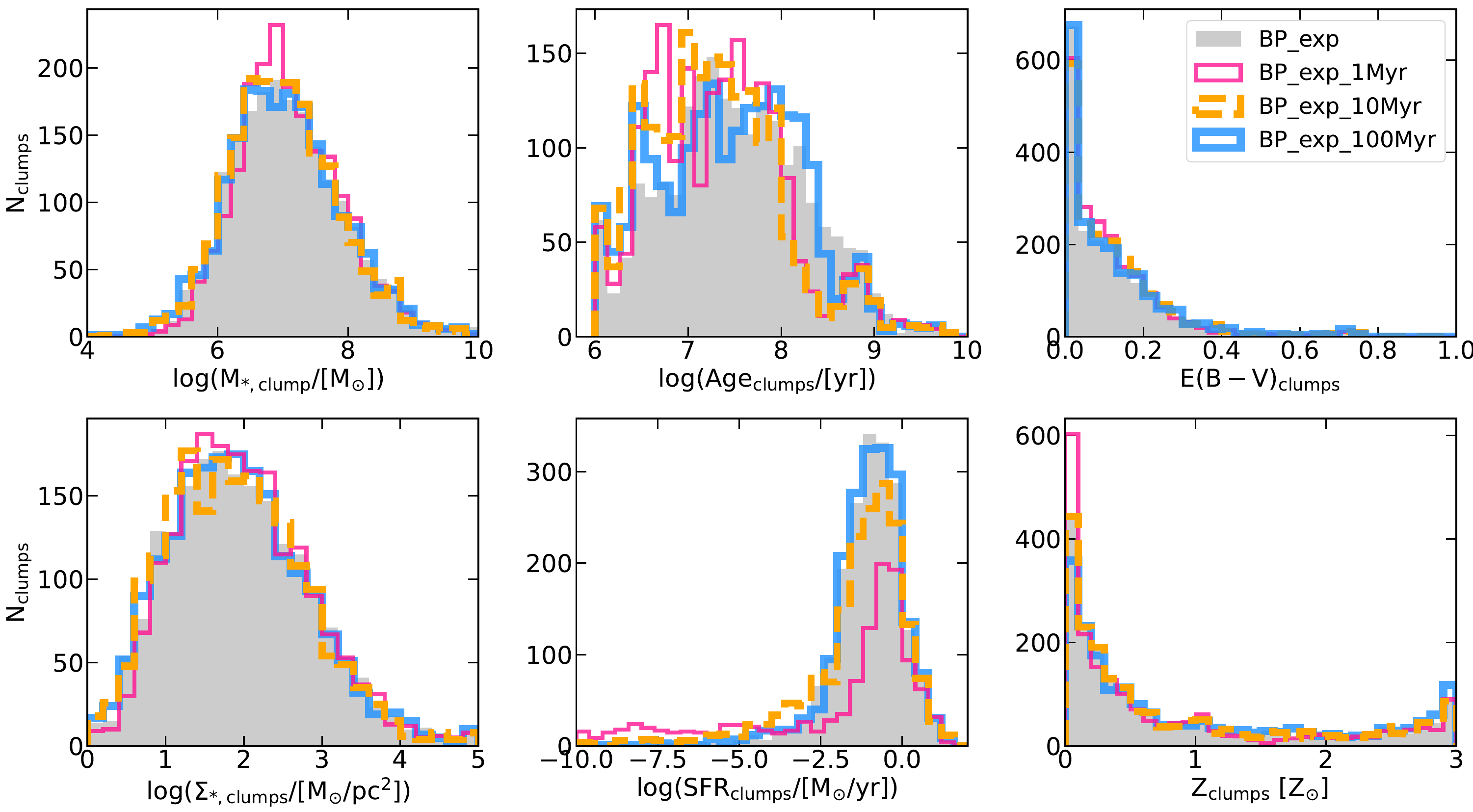}
    \caption{The different clump parameters obtained from SED fitting with four different SFHs: exponential declining with $\tau=1 \ \rm Myr$ (pink), with $\tau=10 \ \rm Myr$ (orange), with $\tau=100 \ \rm Myr$ (blue), and with $\tau$ as a free parameter (grey). The four SFHs produce very similar distributions for the stellar mass ($\rm M_{*}$), the extinction ($\rm E(B-V)$), the stellar surface density ($\Sigma_{*}$), and the metallicity ($\rm Z_{*}$). As expected, the choice of the SFH affects the stellar age of the clumps ($\rm Age$) with the SFH with $\tau=100 \ \rm Myr$ producing significant older ages.  }
    \label{fig:clumps_SFH}
\end{figure*}

\noindent Fig.~\ref{fig:clumps_SFH} presents the distribution of the main properties (stellar mass $\rm M_{*}$, age, extinction $\rm E(B-V)$, stellar surface density $\rm \Sigma_{*,clumps}$, SFR and metallicity $Z$) of the clumps for the 4 assumed SFHs.  We use mass-weighted ages computed by weighting the mass of stars at the time of their formation:
\begin{equation*}
    \rm age=\frac{\int _{0}^{t_{obs}} t\times SFR(t) dt}{\int _{0}^{t_{obs}} SFR(t) dt },
\end{equation*}
where $\rm t_{obs}=t(z_{obs})$ and $SFR(t)$ correspond to the assumed SFH. The mass-weighted age gives an indication of the main epoch of the stellar build-up of the clumps.

\noindent We notice that the global distributions of stellar masses, extinctions, densities and SFRs are very similar among the different models. Only the ages are affected by the choice of duration of the star formation burst, with short SFHs (i.e., with $\rm \tau=10~Myr$ and $\rm \tau = 1~Myr$) producing younger ages (i.e, age$<10$ Myr) and under-populate the age bin between 0.1 and 1 Gyr with respect to fits obtained with longer SFHs. Overall, we observe that the fit with $\tau$ free (BP\_exp) provides a more flexible constraint on the duration of the burst. The latter produces an age range that overlaps with the other distributions, but with significant differences both at very young (below 10 Myr) and older ages ($>$100 Myr). We therefore opt to use the SED fit outputs produced by the later model as reference for the remaining of the analysis. Finally, we notice that of all output parameters, the recovered metallicity seams the most degenerate, as already expected due to the lack of spectral features in the SED that can provide tight constraints.

\noindent The SFH used to reproduce clumps SED do not allow for multiple burst of star formation. Such SFH could produce stellar populations with different formation times, including recent star forming activity, and then give older mass-weighted ages.

\subsection{Lens model and magnification estimations}
\label{sec:lens_model}

At this stage, the resulting clump catalogue includes all the 2,004 clumps that have detections in two reference JWST bands, F150W and/or F200W and/or F277W. In order to estimate the intrinsic properties of the clumps, we use the galaxy cluster lens model to produce magnification maps of each galaxy in the image plane. 

\noindent Our baseline model and methodology to constrain the cluster mass distribution is the Lenstool \citep{Jullo2007} parametric model described in \citet{Mahler2018}, and which includes a combination of cluster-scale and galaxy scale mass components as double pseudo-isothermal elliptical (dPIE- profiles. The new JWST/UNCOVER data have pushed the need for an additional cluster-scale component in the north-west, which we constrain using the same multiple images as \citet{Furtak2023}. Our lens model is able to reproduce the set of 136 multiple images coming from 46 different sources with an rms of 0.76".

\noindent At the location of the clump within the galaxy, we measure the median value from the corresponding magnification map in the region enclosed within the ellipsoidal describing the shape of the clump (using the best-fit parameters $x_{\rm std}$, $\epsilon$, and $\alpha$ as measured on the reference filter). We measure three magnification values for each clump, the total magnification, $\mu$, the radial magnification, $\mu_R$, and the tangential magnification, $\mu_T$ ($\mu= \mu_R \times \mu_T$). The uncertainties introduced by the magnification are the sum in quadrature of two components ($\delta\mu_1$ and $\delta\mu_2$). First, we measure the standard deviation of the magnification values within each clump region, which represents the potential spatial variation of magnification across the size of the clump ($\delta\mu_1$, this value is particularly high for the clumps located very close to the critical lines). Second, we randomly produce 100 magnification maps for each clump, selected from the lenstool MCMC posterior distributions of the lens model. We measure the magnification of each clump on the 100 maps and estimate the 68\% confidence interval on each side of the best value to obtain asymmetrical errors on the magnification ($\delta\mu_{2-}$ and $\delta\mu_{2+}$ for lower and upper error-bars).
The final lower and upper magnification uncertainties of each clump is given by $\delta \mu _{tot+,-}=\sqrt{\delta \mu 1^2 + \delta \mu_{2+,-}^2}$.

\subsection{Magnification corrections}
\label{sec:magnification_corrections}

Intrinsic fluxes are derived by dividing the observed fluxes by these final magnification values. The intrinsic fluxes are converted into AB absolute magnitudes after correcting from the Galactic reddening \citep{SchlaflyFinkbeiner2011} in each filter. 
To derive the intrinsic effective radius, $R_{\rm eff}$, we first estimate the radius of the circle having the same area of the ellipses describing the morphology of the clump, i.e., $R_{\rm cir} = \sqrt{x_{\rm std}\times y_{\rm std}}$. We assume that $R_{\rm cir}$ is the standard deviation of a 2D circular Gaussian, and derive the observed PSF-deconvolved effective radius as $R_{\rm eff,obs}=R_{\rm cir}\times \sqrt{2 ln(2)}$. 

\noindent We consider three cases when measuring the intrinsic effective radius. According to the testing performed by \cite{Messa2019} and \cite{Messa2022}, we assume 0.4 pixel as the minimum resolved Gaussian axis standard deviation (std) in the reference filter. If the clump is resolved along both axes in the image plane, the intrinsic effective radius, $R_{\rm eff}$, is obtained by dividing $R_{\rm eff, obs}$ by $\sqrt{\mu}$. If the clump is not resolved in one direction we divide $R_{\rm eff, obs}$ (with $R_{\rm cir}=x_{\rm std}$ and $x_{\rm std}$ the axis of the resolved direction) by the magnification along the shear direction ($\mu_t$). The underlying assumption in the latter case is that the clump is intrinsically circular and its radius is resolved in the tangential direction \citep[e.g.,][]{Vanzella2017}. The same assumption (circular shape) is made in the third case, i.e., if the clump is unresolved in both directions. In this latter case we use $\mu_t$ to derive the size upper-limit. Finally, the physical sizes in parsecs are derived considering the pixel size of the F150W images (0.02'') and the angular diameter distance of each galaxy. We propagate the magnification in the effective radius errors. Stellar masses and SFR obtained from SED fitting are divided by the total magnification value of each clump to get the intrinsic values and the magnification errors are propagated.


\section{Results}
\label{sec:results}

\subsection{Clump luminosity-size relation as a function of redshift}
\label{sec:Mag_Reff}

\begin{figure}
	\includegraphics[width=9cm]{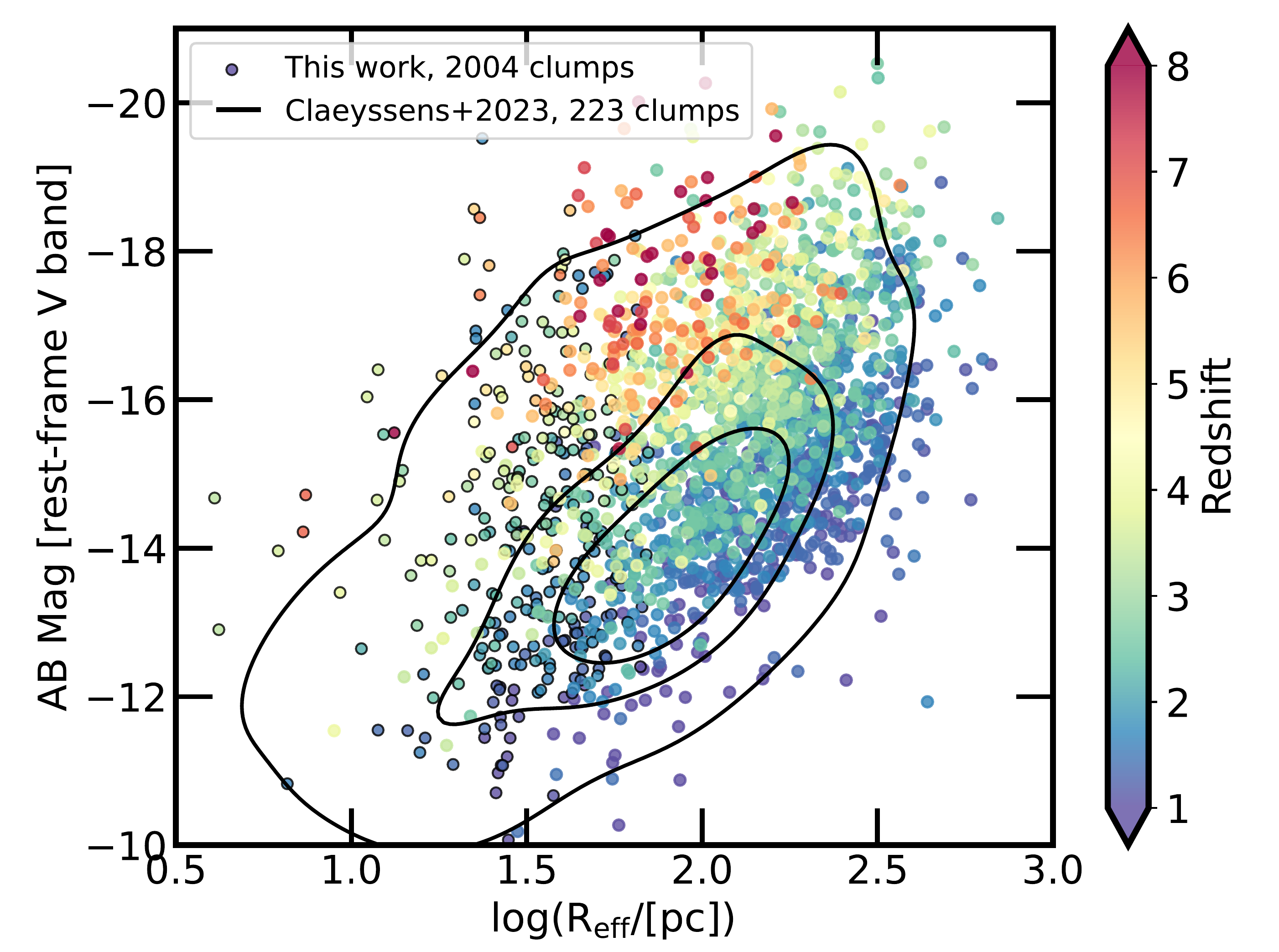}
    \caption{Clumps effective radius $\rm R_{eff}$ plotted versus V band rest-frame absolute AB magnitude. The points are color-coded by redshift, the points circled in black are upper limits in size. The black contours enclose 90\%, 50\% and 25\% of the 223 clumps sample presented in \citetalias{Claeyssens2023}. }
    \label{fig:Mag_Reff}
\end{figure}

Numerous clumps studies have used the luminosity (or SFR) versus size relation as a tool to compare physical conditions of clumps as a function of redshift. Using a compilation of several literature samples including magnified stellar clumps at $z<3$ as well as star-forming regions in typical main-sequence galaxies in the local universe, \citet{Livermore2015} measured that higher surface brightness clumps are found in higher redshift galaxies. This trend has been linked to the fact that clumps form out of fragmentation in galaxies that have higher gas fractions for increasing redshift. Differently to initial works, we are now able to look at this relation using V-band rest-frame properties, thus being less affected by selection biases introduced by UV wavelengths. In Fig.~\ref{fig:Mag_Reff} we show the V-band rest-frame AB absolute magnitude of the clumps as a function of their effective radius and color-coded on their redshift. The magnitude is estimated from the F090W filter at $0.7<z<1.5$, F115W at $1.5<z<2.5$, F150W at $2.5<z<3.5$, F200W at $3.5<z<5.5$ and F277W at $5.5<z<10$. As a comparison we overplot the contours that encompass the 90\%, 50\%, and 25\% of the sample of 223 clumps detected in the JWST/NIRCam observations of the lensing cluster SMACS0723 (\citetalias{Claeyssens2023}). The A2744 sample probes a similar parameter space as found in  SMACS0723 sample. However, we detect only 41 clumps with $\rm R_{eff}<20 \ pc$ in A2744 (i.e., 2.1\%), while 35 have been studied in the SMACS0723 sample (i.e., 15\%). This effect is due to a combination of different magnification and detection limits in the two datasets. We observe a clear trend with redshift consistent with the findings of \citet{Livermore2015} suggesting that clumps can reach higher stellar surface densities at higher redshift. However, this distribution is also strongly affected by completeness effect at the faint surface-brightness end due to cosmological dimming at higher redshift. This effect is also visible in Fig.~\ref{fig:Magnification} where we see that the majority of clumps have magnifications below 10 and that for similar magnifications the fainter clumps are detected only in the lower redshift bin. Hence, while clumps at increasing redshift reach higher SFR densities reflecting the changing physical conditions of the host galaxies where they form, it remains unclear whether the lower densities clumps are not forming or are simply missed.

\subsection{The effect of magnification on derived clump physical properties}
\label{sec:magnification}

For the rest of the paper, we focus only on the clumps and galaxies with $z<5.5$, . The higher redshift clumps ($z>5.5$), located in the reionisation era, will be studied in an upcoming paper. The $z<5.5$ subsample contains 377 individual galaxies (385 images) and 1751 (1787 with multiple images) clumps. Among these clumps, 34 have robust detection (i.e., with a SNR $>3$) in less than 4 NIRCam broad-band filters; as a consequence, their SEDs are poorly constrained and we exclude those from the rest of the analysis. We divide the sample in four redshift bins encompassing the cosmic noon: $0.7\leq z<1.5$, $1.5 \leq z<2.5$, $2.5\leq z<3.5$  and $3.5\leq z<5.5$  containing 62, 121, 101, and 93 galaxies, respectively (with 410, 630, 435 and 276 clumps, respectively). We show the clumps properties using the reference exponential decline SFH with $\tau=10$ Myr. As discussed above this choice does not strongly affect any of the physical properties except the age distributions which we will discuss comparing different SFH assumptions (cf Fig.~\ref{fig:clumps_SFH}).

\begin{figure*}
	\includegraphics[width=18cm]{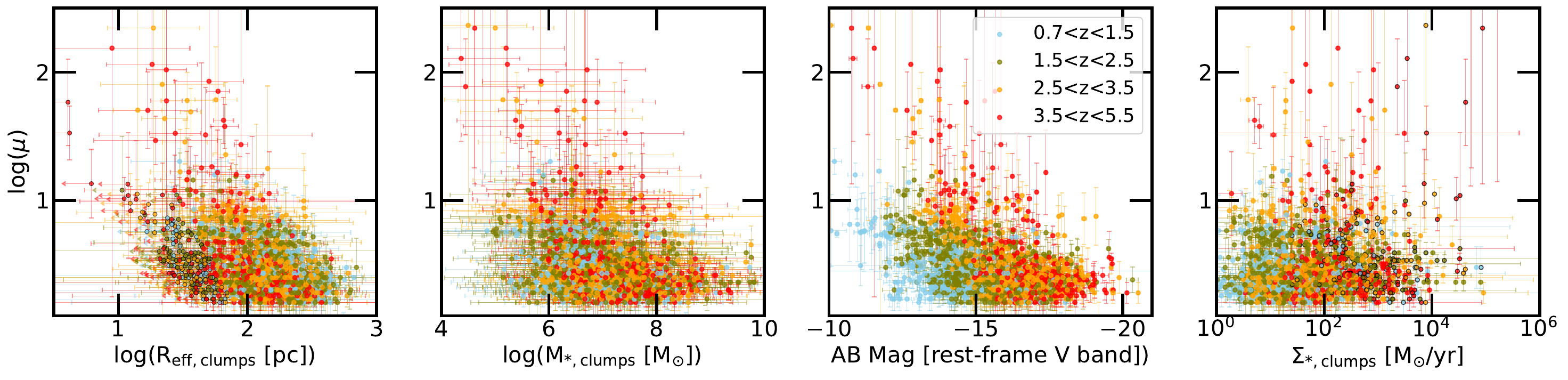}
    \caption{From left to right: lensing magnification as a function of the clumps effective radius, stellar mass, AB absolute magnitude and stellar surface density ($\rm \Sigma_{*,clumps}$). The points are color-coded in the four redshift bins. The points circles in black in the first and third panels are upper limits in size and lower limits in density, respectively.}
    \label{fig:Magnification}
\end{figure*}

\noindent Fig.~\ref{fig:Magnification} shows the average magnification of the clumps as a function of their effective radius, stellar mass, absolute AB magnitude and stellar surface density. 
The highest magnifications ($\mu \sim$ 100) are achieved in proximity of the critical lines and are also the values with the largest uncertainties and mostly belong to the $2.5<z<5.5$ redshift bins. We found that 12\% of the clumps are unresolved, i.e., we can only estimate upper limits which range between 10 and 60 pc (the effective physical resolution impact on the clumps properties is discussed in details in the Sect.~\ref{sec:resolution}). These systems line-up forming a sequence in the first panel of Fig.~\ref{fig:Magnification}. 
We observe a strong impact of the magnification on the probed distribution of size and magnitude with redshift. At $z>3.5$, stellar mass $<10^6 \rm \ M_{\odot}$ are only detected at magnification higher than 80, while they can be detected at lower redshift at $\mu=2-80$. 

\noindent The quantity the most affected by the magnification is the absolute rest-frame V-band magnitude (third panel in Fig.~\ref{fig:Magnification}). For similar magnification ranges, we observe a significant shift in the faintest magnitudes probed in the four redshift bins. The minimum magnification-magnitude sequence in each redshift bins illustrate the completeness of our sample. Our ability to detect faint clumps (i.e, AB Mag $>-15$) at $z>2.5$ depends on the number of clumps magnified by at least $\mu>100$. 

\noindent The last panel of Fig.~\ref{fig:Magnification} shows the stellar surface density of the clumps. This quantity depends from both the intrinsic size and stellar mass of the clump and thus is independent of the magnification, but sensitive to surface brightness dimming effects. Indeed, the distributions of stellar surface density versus magnification do not show any correlation between the two parameters. Very high magnifications probe a wide range of stellar surface densities, even at $z>2.5$. However, we detect very few low densities (i.e., $\rm \Sigma_{*,clumps}<10^{1.5} \ M_{\odot}/pc^2$) clumps at $z>3.5$ and they all present a high magnification ($\mu>80$) likely because of surface brightness dimming effects. 

\subsection{Galaxy properties across cosmic time}
\label{sec:galaxies_redshift}

\begin{figure}
	\includegraphics[width=9cm]{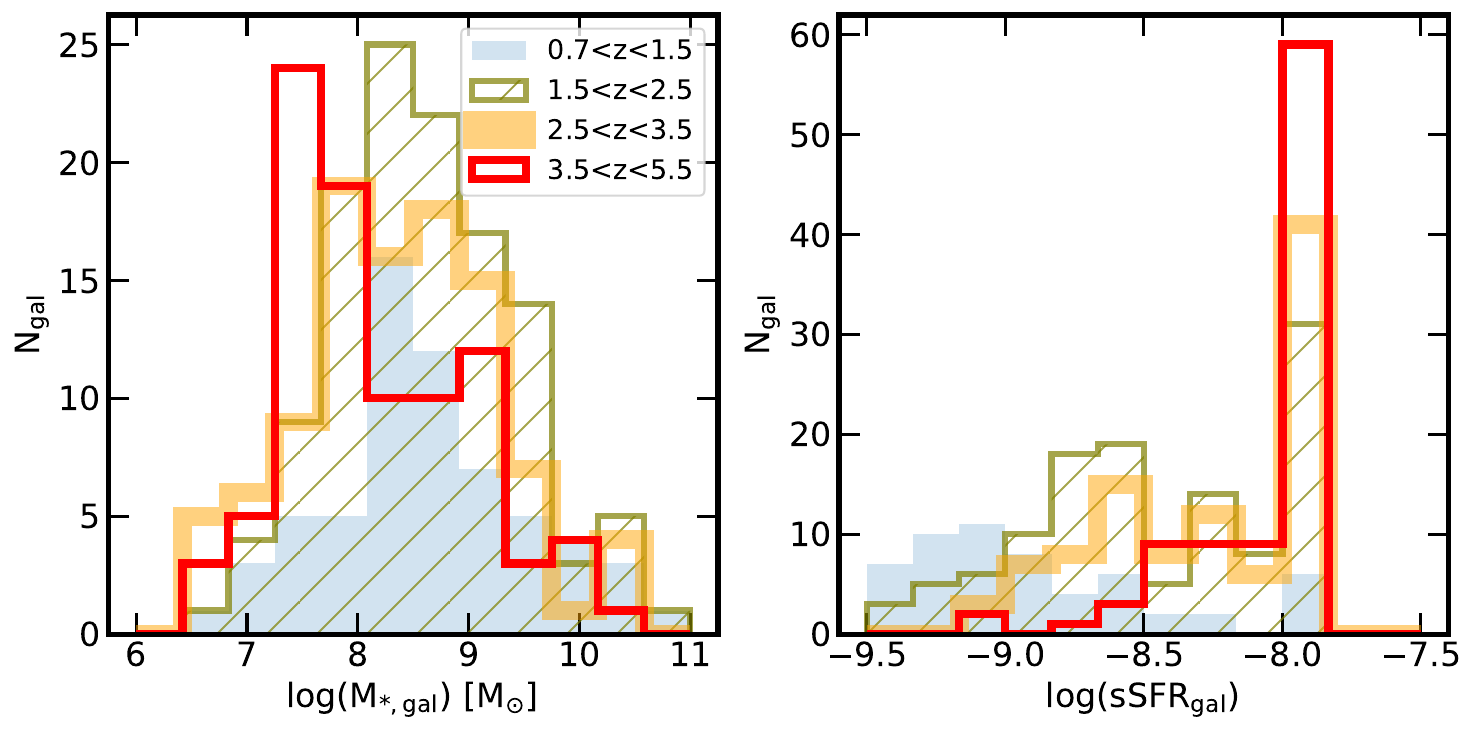}
    \caption{Distribution of the galactic stellar mass ($\rm M_{*,gal}$) and specific SFR ($\rm sSFR_{gal}$) in four redshift bins. The sSFR peak around $-8$ corresponds to high-redshift galaxies which formed all their mass over the last 100Myr.}
    \label{fig:galaxies_redshift}
\end{figure}

We investigate the variations of integrated galaxy properties across cosmic time, through the four redshift bins.
Fig.~\ref{fig:galaxies_redshift} presents the galaxy stellar mass ($\rm M_{*,gal}$) and specific SFR ($\rm sSFR_{gal}$) defined as the SFR (estimated over the last 100 Myr in our case) divided by the stellar mass. The sSFR compares the current star formation to the averaged past one, a high value of sSFR indicates that the galaxy is still quite active in star formation. Stellar mass ranges from a few times $10^6$ to $10^{11} \ \rm M_{\odot}$. The highest redshift bin contains a significantly lower quantity of high-mass galaxies (with only 17 ($\sim 18\%$) galaxies with $\rm M_{*,gal}>10^9 \ M_{\odot}$) and a large number of low-mass galaxies (with 51 ($\sim 55\%$) galaxies with $\rm M_{*,gal}<10^8 \ M_{\odot}$). The average distribution of sSFR is strongly evolving with redshift. Galaxies at $z<1.5$ have mostly very low sSFR values with 77\% objects with $\rm log(sSFR)<-8$, while galaxies at $3.5<z<5.5$ have very high sSFR values with 88\% galaxies with $\rm log(sSFR)>-8$. The peak of sSFR around $-8$ is produced by galaxies having formed all their mass in the last 100 Myr, which represents 60\% of the $3.5 \leq z < 5.5$ galaxies. 

\subsection{Clump properties across cosmic time}
\label{sec:clumps_redshift}

\begin{figure*}
	\includegraphics[width=18cm]{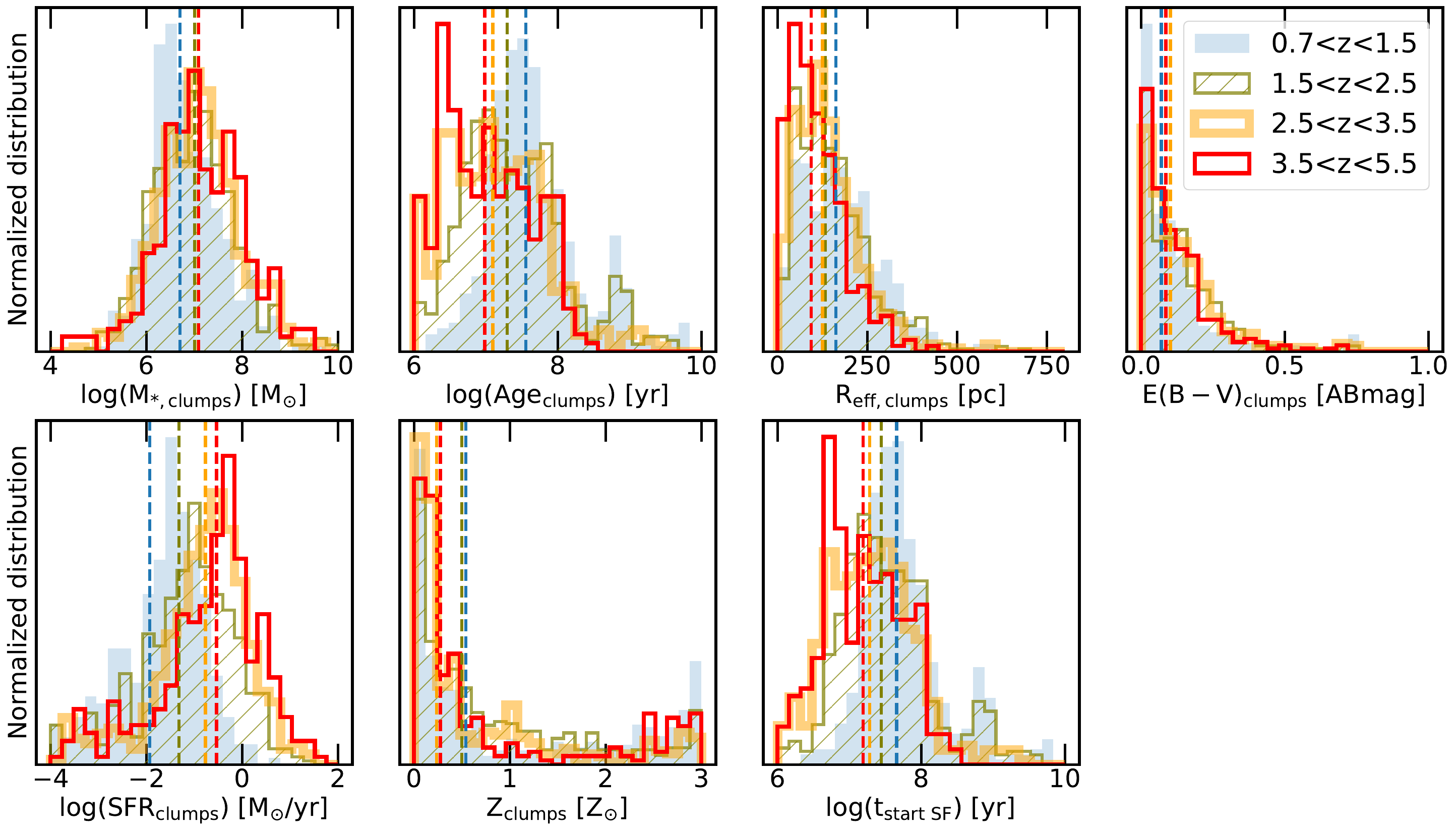}
    \caption{The different clump parameters obtained from SED fitting (stellar mass $\rm M_{*,clumps}$, age ($\rm Age_{clumps}$), extinction ($\rm E(B-V)$), SFR ($\rm SFR_{clumps}$), and metallicity ($\rm Z_{clumps}$)) with the reference SFH (exponential decline with $\tau=10$ Myr) and the clumps effective radius ($\rm R_{eff, clumps}$), shown for four different redshift bins. The median of each distribution is represented with the dashed vertical line.}
    \label{fig:clumps_redshift}
\end{figure*}

We investigate in Fig.~\ref{fig:clumps_redshift} the redshift variations of the clump properties, using the same redshift bins introduced in the previous section. With the exception of size, mass and SFR, the intrinsic clump properties estimation does not depend on the lens magnification estimation robustness and uncertainties. 

\noindent We do not observe significant variations in the extinction ($\rm E(B-V)_{\rm clumps}$), as between $78$ and $85\%$ of clumps have very low extinction values $\rm E(B-V)<0.2 \ mag$, independently of their redshift. The stellar mass distribution is also relatively stable with redshift, with a slight increase of the median stellar mass at higher redshift (from $\rm M_{*,clumps}=10^{6.7} \ M_{\odot}$ at $0.7<z<1.5$ to $\rm M_{*,clumps}=10^{7.1} \ M_{\odot}$ at $3.5 \leq z <5.5$). The distributions of the metallicity is stable across redshift with most of the clumps having a metallicity lower than $\rm 50\% \ Z_{\odot}$. 

\noindent We measure significant variations in age ($\rm Age_{\rm clumps}$), size ($\rm R_{\rm eff, clumps}$), SFR ($\rm SFR_{\rm clumps}$) with redshift.

\subsubsection{Size}

The sizes of the clumps range from $1$ to $700$ pc. In total, 75\% of the clumps are smaller than 200 pc and 37\% smaller than 100 pc. While we observe a wide range of sizes in every redshift bin, the median values evolve with redshift from 162 pc at $0.7<z<1.5$ to 93 pc at $3.5<z<5.5$. The size distribution is very stable between $z=1.5$ and $z=3.5$ with the two middle redshift bins having similar median values at 125 pc. This evolution could be affected by both the magnification (cf Fig.~\ref{fig:Magnification}, high-redshift bin has the largest magnifications) and by the redshift evolution of the angular diameter distance affecting the spatial resolution.

\subsubsection{SFR}

We measure a significant increase of the SFR with increasing redshift with the median values going from $\rm SFR=10^{-1.9} \ M_{\odot}/yr$ at $0.7<z<1.5$ to $\rm SFR=10^{-0.5} \ M_{\odot}/yr$ at $3.5<z<5.5$. Higher redshift clumps present, on average, a higher SFR value, which could be due also to younger stellar populations. This trend is consistent with other clump studies, which reported higher SFR and stellar surface densities with increasing redshift (see \citealt{Livermore2015, Messa2022, Messa2024} for HST clump studies and \citetalias{Claeyssens2023} for a JWST clump study).
The intrinsic SFR estimation depends on the lens magnification, so the robustness of individual measurements could be affected by the magnification uncertainties, however the stellar surface and SFR densities estimations are independent of the magnification as the latter factor enters in both size and mass (or SFR). Therefore, the stellar surface density and the SFR density is a more robust parameter in the study of clump properties in lensed galaxies. However, the stellar and SFR density distributions can be affected by completeness effect, especially at high redshift where it becomes more difficult to detect faint structures due to  cosmological surface brightness dimming. We discuss the clump stellar surface density and SFR evolution in Sect.~\ref{sec:redshift_evolution}.

\subsubsection{Age}

The maximum clump age measured is 6 Gyr at $z<1.5$ and the minimum is 1.0 Myr. The age estimate from UV-optical SED fitting can be very uncertain and strongly affected by degeneracies, mainly with dust and metallicity. As discussed above we get equally good SED fits in spite of the assumed SFHs. However, once the SFH is fixed, the access to 20 JWST/NIRCam filters, sampling the UV-optical of stellar continuum and nebular emission lines (depending on redshift), provides robust ages. We measure a global median age uncertainty of 0.31 dex with the first and third quartiles at 0.27 dex and 0.37 dex, respectively; these values are significantly better than the ones achieved in \citetalias{Claeyssens2023} with observations from only 6 BB NIRCam filters. This improvement is mainly due to the addition of 2 broad-band filters (F070W and F090W) and 12 medium-band filters in the SED fit thanks to the last observation program of A2744 (see Sect.~\ref{sec:JWST_obs}). We compared the age uncertainties obtained with the same SED fits with and without the F070W, F090W, and all the medium-band filters, and found a strong improvement in the age uncertainty distribution. Without the additional filters, we obtained a median age uncertainty of $\rm log(Age)=0.40$, with the first and third quartiles at $\rm log(Age)=0.35$ and $\rm log(Age)=0.45$, respectively (cf Sect.~\ref{sec:appendix} for more details on these tests).

\noindent The clump mass-weighted age distribution evolves significantly with redshift: high-redshift clumps are, on average, younger than clumps within $z<1.5$ galaxies. this evolution is present in spite of the assumed SFH. 
For the reference SFH assumption ($\tau=10$ Myr), we find that the median age evolves from 36 Myr at $0.7<z<1.5$ to 9 Myr at $3.5<z<5.5$. We do not find any clump older than 320 Myr at $3.5<z<5.5$. At $z<3.5$, we measure 128 clumps older than 300 Myr (9\%). The global trends with clumps age remain consistent with the different SFH presented in Sect.\ref{sec:clumps_SED} and Fig.\ref{fig:clumps_SFH}, although we notice that at $z<3.5$, 11 \% of the clumps show ages older than 300 Myr.

\noindent This shows that a fraction of clumps can survive for longer time than what has been predicted in some simulations of clumps in distant galaxies (\citealt{Genel2012,Tamburello2015}). The existence of long-lived clumps was also predicted by several simulations under specific conditions of gas and stellar densities and different types of feedback prescriptions (\citealt{Perret2014,Bournaud2014,Ceverino2014,Fensch2017,Calura2022,Ceverino2023}). 

\noindent We observe a significant variation of the age of the starting point of the star formation burst ($\rm t_{start \ SF}$). At all redshift, with a short SFH  (i.e., an exponential decline of $\tau=10$ Myr) the clumps present young stellar population, with a start formation having started typically less than 100 Myr ago. We discuss further the physical implications of the ages distribution in Sect.~\ref{sec:migration}.


\subsection{Host galaxies and clumps properties}
\label{sec:host_galaxies}

In this section, we describe the link between the clump properties and their host galaxies. In particular, we focus on the global galactic physical properties that are more robustly derived, e.g., stellar mass, SFR and sSFR.

\begin{figure*}
	\includegraphics[width=18cm]{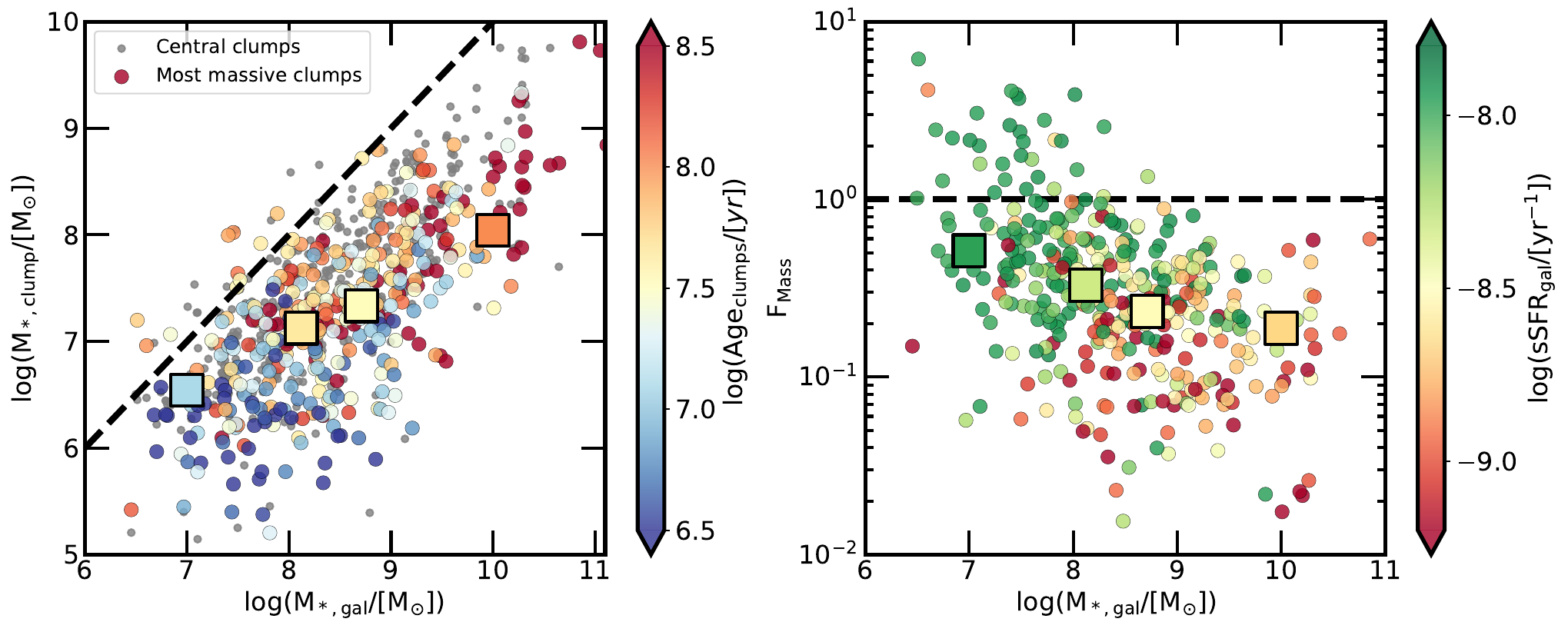}
    \caption{Left: Stellar mass ($\rm M_{*,clumps}$) of the central clumps (grey circles) and of the most massive clumps (non-central, coloured stars) as a function of the host galaxy stellar mass. The black dashed line represents the 1:1 line.
    The points are colour-coded by the age of the most massive clump per galaxy, and the median values by the median value of these ages in each bin. The squares represent the median values measured in four $\rm M_{*,gal}$ bins containing the same number of galaxies. Right: Clump mass fraction $\rm F_{Mass}$ (defined as the clumps mass divided my the host galaxy mass) as a function of the host galaxy stellar mass $\rm M_{*,gal}$. The points are colour-coded by the specific SFR of the host galaxy ($\rm sSFR_{gal}$). The squares represent the median values measured in four $\rm M_{*,gal}$ bins containing the same number of galaxies. The black dashed line represents $\rm F_{Mass}=1$.}
    \label{fig:host_galaxies}
\end{figure*}

\noindent In order to identify the central clump of each galaxy, we projected the image of each galaxy (in the reference filter) in the source plane to recover its intrinsic global morphology. This source reconstruction method is only based on the direct inversion of the lensing equation for each pixel and does not include PSF deconvolution and noise subtraction. Therefore, as the PSF is projected in the source plane, it  does not allow to perform resolved measurements within the reconstructed sources, such as the clump photometry. However, given that the NIRCam PSF is very sharp compared to the observed extent of the galaxies (see Table \ref{tab:table_observations} for the values in each filter), it does not affect the global morphology of the reconstructed source. This type of reconstruction is accurate enough to identify the central clump, which is defined as the closest clump to the light barycenter. This approach allow us to locate the barycenter on top of the proto-bulge and bulge within evolved galaxies (mostly at $z<2.5$), however, identifying a central clumps within young and strongly irregular galaxies is more complicated. 

\noindent The identification of the central clump and in general of the projected distance of the clumps from the centre is particularly useful when we address the contamination of proto-bulges in our analyses, or look for trends as a function of the galactocentric distances.

\noindent Fig.~\ref{fig:host_galaxies} (left panel) presents the maximum clump stellar mass (when excluding the most central clump), as well as the central clump stellar mass measured in each galaxy as a function of the $\rm M_{*,gal}$. 
We measure a significant correlation between the maximum clump mass and central clump mass with the host galaxy mass. Indeed, we measure a Pearson coefficient of $\rm r=0.80$ with a p-value $\rm p_0<10^{-88}$ between $\rm log(M_{*,clumps})$ and $\rm log(M_{*,gal})$ for the central clumps, and  $\rm r=0.72$ with a p-value $\rm p_0<10^{-60}$ for the rest of the clumps. We divided the sample in four bins based on the host galaxy mass ($\rm log(M_{*,gal}/[M_{\odot}])<7.87$, $\rm 7.87<log(M_{*,gal}/[M_{\odot}])<8.37$ , $\rm 8.37<log(M_{*,gal}/[M_{\odot}])<9.05$, and $\rm log(M_{*,gal}/[M_{\odot}])>9.05$), containing between 92 and 96 galaxies each. 
We measure the median maximum clump stellar mass and the central clump stellar mass in each bin, and confirm a clear increase with the host galaxy mass. We measure, from low-mass to high-mass galaxies, median values of $\rm log(M_{*,clump}/[M_{\odot}])=6.9$, $7.4$, $7.9$, and $8.7$ for the central clumps, and $\rm log(M_{*,clump}/[M_{\odot}])=6.4$, $7.0$, $7.3$, and $8.0$ for the most massive other clumps. Overall, for any galaxy mass bin, the central clumps are, on average, more massive than any other clumps.

\noindent We measure a second correlation between the stellar mass and age of the clumps as a function of galaxy mass, with the most massive clumps on average, older (see the color-coding of the left panel in Fig.~\ref{fig:host_galaxies}). The respective ages of the median most massive clumps are $\rm Age=12$ Myr, $\rm Age=39$ Myr, $\rm Age=32$ Myr, and $\rm Age=95$ Myr from low-mass to high-mass galaxies. The most massive structures formed in very massive galaxies contain older stellar populations, on average in agreement with the physical properties distributions discussed above. 

\noindent Fig.~\ref{fig:host_galaxies} (right panel) presents the clump mass fraction ($\rm F_{Mass}$) as a function of the galactic mass colour-coded by the galactic sSFR. $\rm F_{Mass}$ is defined as the sum of the mass of all the clumps (including the central clumps) within each galaxy divided by the host galactic mass, and represents the fraction of stellar mass located in the clumps. We measure a significant anti-correlation between $\rm F_{Mass}$ and $\rm log(M_{*,gal})$ with a Pearson coefficient of $\rm r=-0.30$ and a p-value $p_0<10^{-10}$. We measure the median mass fractions in the same four galactic mass bins as previously, which confirm a clear decrease of the mass fraction with increasing host galaxy mass, with median values of $\rm F_{Mass}=0.53$, $0.35$, $0.23$, and $0.19$ from low-mass to high-mass galaxies. 
A correlation between $\rm F_{Mass}$ and the sSFR of the host galaxy is also observed. Galaxies with the higher $\rm F_{\rm Mass}$ present significantly higher sSFR, with median values of $\rm log(sSFR/yr^{-1})=-7.9$, $-8.3$, $-8.5$, and $-8.7$. 

\noindent These results tend to show that the most massive galaxies present a lower mass fraction in clumps, associated with a lower sSFR value. On the other hand, lower mass and less evolved galaxies present a higher mass fraction in clumps and a very high sSFR. The latter trend can be understood if a significant part of the star formation is taking place in clumps. This trend is also linked to    redshift evolution, as we observe a higher fraction of low-mass and high sSFR galaxies at higher redshift (cf Fig.~\ref{fig:galaxies_redshift})

\noindent In Fig.~\ref{fig:host_galaxies} we observe that in a few cases (29 galaxies) the (total) mass of the (central) most massive clump can be larger than the mass recovered for the host galaxy.
When performing SED fitting based on the total galaxy photometry, using simple SFH assumptions, we are averaging the contributions of the individual clumps and diffuse light from the galaxy. 
\citealt{Gimenez2023,gimenezarteaga2024_arxiv}, using JWST/NIRCam data, have shown that SED fitting on integrated photometry could yield to younger ages and lower stellar masses when compared to spatially resolved spectral energy distribution fitting. This effect is similarly observed for a small fraction of our targets.


\subsection{Galactocentric evolution of clumps properties within galaxies}
\label{sec:galactocentric_trends}

\begin{figure*}
	\includegraphics[width=18cm]{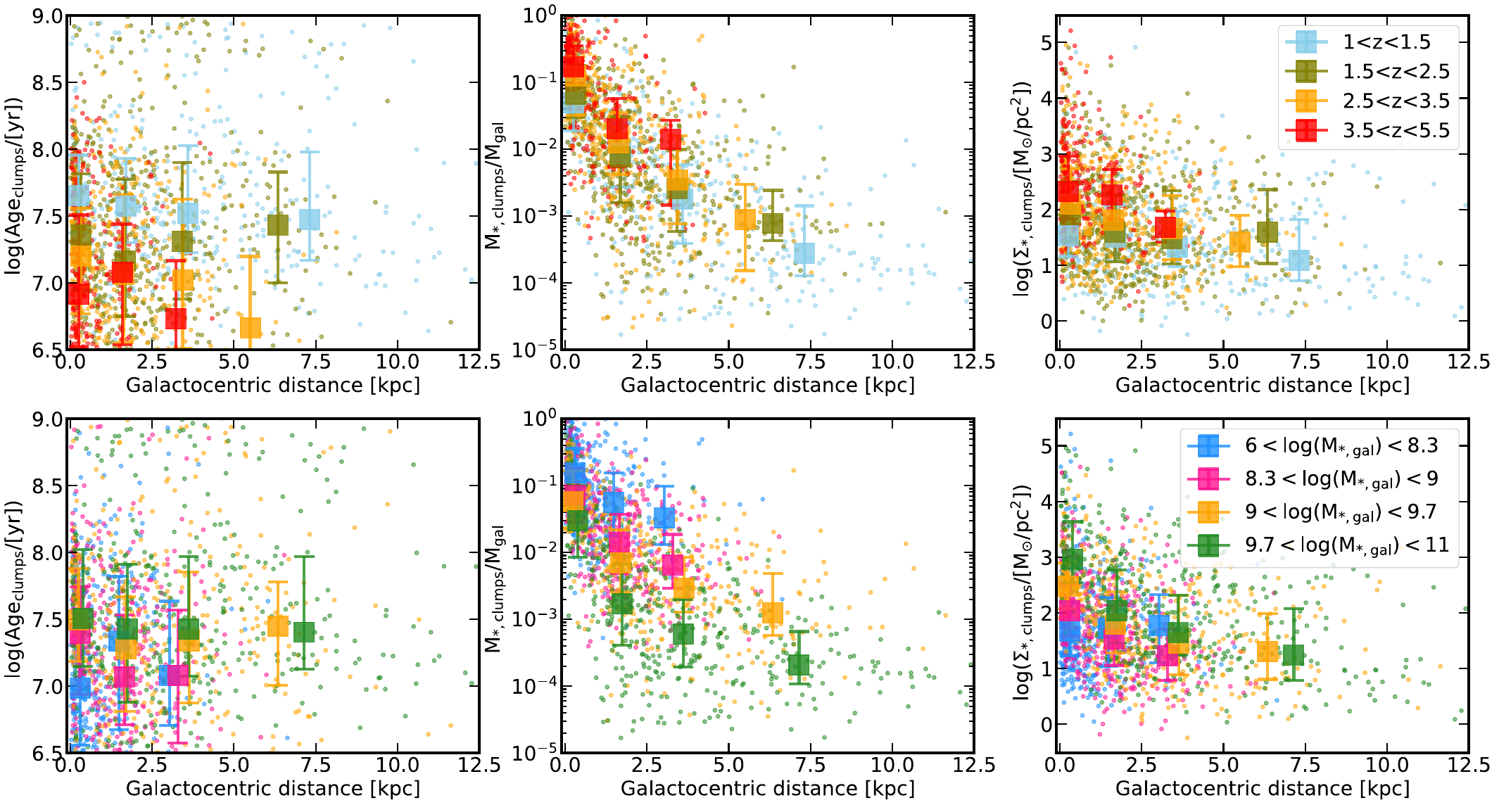}
    \caption{Clump properties as a function of their galactocentric distance in the host galaxies. Top row: mass-weighted ages ($\rm Age_{clumps}$), stellar mass fraction ($\rm M_{*,clumps}/M_{*,gal})$), and stellar surface densities ($\rm \Sigma_{*,clumps}$) of the clumps as a function of their galactocentric distance. The points are colour-coded according to the redshift, divided in four redshift bins. The squares show the median, first quartile and third quartile measured in four galactocentric distance bins defined as $0\leq z<d<1$ kpc, $1 \leq d<2.5$ kpc, $2.5 \leq d<5$ kpc, and $5 \leq d<12.5$ kpc. For galaxies with $z>3.5$, we consider only three galactocentric distance bins ($0 \leq z<d<1$ kpc, $1 \leq d<2.5$ kpc, $2.5 \leq d<12.5$ kpc) as the galaxies are much smaller. Bottom row: Same clump quantities as a function of their galactocentric distance, colour-coded according to their host galaxy mass. We divided the sample in four galactic mass bins containing the same number of clumps. The squares show the median, first quartile and third quartile measured in the same four galactocentric distance bins.  }
    \label{fig:migration}
\end{figure*}

Several works in the literature have analysed and discussed galactocentric trends of the clumps physical properties  (\citealt{Dekel2009, Ceverino2010, Krumholz2010, Cacciato2012,Forbes2012,Guo2018,Krumholz2018,Dekel2020, Dekel2021,Kalita2024}). We use the source reconstruction procedure presented in Sect.~\ref{sec:host_galaxies} to project the position of each clump in the source plane and measure their distance from the barycenter of light. In Fig.~\ref{fig:migration} we present the clump stellar masses, ages, and stellar surface densities with respect to their galactocentric distance in the four established redshift bins (see above) and in four host galaxy mass bins. These latter have been chosen in the way to contain similar number of clumps: $\rm 6<log(M_{*,gal}/[M_{\odot}])<8.1$ (442 clumps from 166 galaxies), $\rm 8.3<log(M_{*,gal}/[M_{\odot}])<9$ (438 clumps from 110 galaxies), $\rm 9<log(M_{*,gal}/[M_{\odot}])<9.7$ (426 clumps from 65 galaxies), and $\rm 9.7<log(M_{*,gal}/[M_{\odot}])<11$ (442 clumps from 36 galaxies). 

\noindent We do not observe any age gradient with galactocentric distance, for none of the redshift nor galactic stellar mass bin, the median age values are stable with the galactocentric distance. This conclusion do not change is we use different SFH for the clump analysis.  However, we measure that globally massive galaxies host older clumps, while low-mass galaxies host younger clumps (in agreement with the distributions seen in  Fig.~\ref{fig:host_galaxies}). We measure a wide range of clump ages at any distance from the galactic center. On the other hand, we notice a redshift evolution of the median clump ages, with clumps being, typically, older in lower redshift galaxies; this result is consistent with the redshift evolution of the clump ages presented in Fig.~\ref{fig:clumps_redshift}. 

\noindent We observe a strong correlation between the individual clump stellar mass fraction (defined as the individual clump mass divided by the host galaxy mass) and the distance from the centre. The individual clump stellar mass fraction becomes higher towards the centre of the galaxies for all redshift and galactic mass bins. The median individual mass fraction values of clumps at $<1$ kpc increase with increasing redshift, ranging from  $5^{+7}_{-3}\%$ at $0.7<z<1.5$ to $18^{+23}_{-1}\%$ at $3.5<z<5.5$ while the median individual mass fraction values of the most outskirts clumps (at $>5$ kpc) range from $0.01^{+0.2}_{-0.02}\%$ at $0.7<z<1.5$ to $0.07^{+0.2}_{-0.05}\%$ at $2.5<z<3.5$. Galaxies at $3.5<z<5.5$ are on average smaller, the median individual mass fraction of the clumps at $>2.5$ kpc is $1.6^{+2.9}_{-1}\%$. 
On average, at all redshift and host galaxy mass, clumps located within 1 kpc from the galactic center represent a median mass around 10\% of the host galaxy mass, while the most outskirt clumps at $>5$ kpc from the galactic center, represent, on average, less than 0.1\% of the galactic mass. The trend is stronger for high-mass galaxies, where outskirt clumps at $>5$ kpc represent a very low mass fraction lower than  0.01\% of the galactic mass. The median values of clumps close to the centers (at $<1$ kpc) increase with decreasing galactic mass, ranging from  $3^{+10}_{-2}\%$ at $\rm M_{*,gal}>10^{9.7} \ M_{\odot}$ to $16^{+19}_{-9}\%$ at $\rm M_{*,gal}<10^{8.3} \ M_{\odot}$. The median outskirt clump values (at $>1$ kpc) increase strongly with decreasing host galaxy mass. This result is consistent with the total mass fraction presented in Fig.~\ref{fig:host_galaxies}.

\noindent The clump surface density is stable with the galactocentric distance in the four redshift bins while higher redshift clumps are one average denser (the redshift evolution of the clumps density is detailed in the Sect.~\ref{sec:redshift_evolution}). However, we observe a significant trend between the clumps stellar mass densities and the galactocentric distance within high-mass galaxies ($\rm M_{*,gal}>10^9 \ M_{\odot}$). Indeed, we observe a strong increase of the clump stellar surface density towards the center of the galaxy, clumps close to the centers are on average denser than outskirt clumps. We do not observe such trend in the lower mass galaxies ($\rm M_{*,gal}<10^9 \ M_{\odot}$) which are also significantly smaller (i.e., with clumps located between 0 and 3 kpc from the center).

\section{Discussion}
\label{sec:discussion}

We built a sample of 1956 individual magnified clumps hosted in 476 individual galaxies (2004 clumps from 484 galaxy images in total) at $0.7<z<10$ from the JWST/NIRCam observations of the lensing cluster A2744. The lens magnification, final values from the last lens model, ranging from $\mu=1.8$ to $\mu=100$, allows to probe faint and compact structures down to $<10$ pc in effective radius and $\rm 10^5 \ M_{\odot}$ in stellar mass.

\noindent We have selected galaxies based on the availability of a robust redshift measurement (spectroscopic or photometric, see Sect.~\ref{sec:galaxy_selection}), meaning that the sample is not complete in terms of a luminosity or mass limit. On the other hand, lensing magnification enables us to reach fainter and higher physical resolutions not necessarily on the brightest or most star-forming galaxies, enabling us to sample galaxies that sit in the main star-forming sequence of galaxies at a given redshift.

\noindent Our sample represents the largest clump sample observed with JWST/NIRCam to date, and is almost ten times larger than the first JWST sample built from the early-release images of the lensing cluster SMACS0723 (\citetalias{Claeyssens2023}, 223 clumps from 18 individual galaxies at $1<z<8.5$). 

\noindent In this study, we focus on the analysis of the clump physical properties of galaxies with redshift $<$ 5.5. 
The redshift range has been selected to probe how star formation operates in galaxies that are rapidly assembling their stellar mass and reach their peak formation at cosmic noon following then by a slow decline of their star formation activity.
Thus, our sample represents a valuable opportunity to study the type of stellar clumps formed in distant galaxies, their evolution across cosmic time, as well as their evolution within their host galaxies.

\subsection{Impact of the effective spatial resolution}
\label{sec:resolution}

\noindent As we aim to break down the structures of star formation to the smallest scale possible, the physical resolution reached has a strong impact on the final clump population detected and characterised. Indeed, the very first studies of clumpy galaxies from HST non-lensed galaxies at $z<2.5$ were claiming the existence of kpc-scale structures shaping the morphology of distant galaxies (\citealt{Elmegreen2007, Elmegreen2005, Guo2011,Guo2012}. However, HST studies of lensed high-redshift galaxies, with magnification factors reaching $\mu=100$ in the most magnified cases, showed that these kpc-scale structures are systematically resolved into clumps of smaller radii (100s of pc), when zooming to smaller scales \citep{DZ2017, Johnson2017, Mestric2022, Messa2022,Messa2024}. Observations of multiple images for some of these galaxies have revealed that the difference of physical resolution reached in each image (due to the difference of magnification produced by the lensing effect) can strongly affect the number of clumps detected \citep[e.g.][]{Johnson2017,Cava2018,Mestric2022, Messa2022, Claeyssens2023}. Two effects are observed in these cases: fainter isolated clumps can be detected in the most magnified image thanks to the magnification boost in flux, and massive clumps which are roughly resolved in the less magnified images are broken down to multiple structures when the shear increases. For instance, in \citetalias{Claeyssens2023}, the counter-image of the Sparkler galaxy at $z=1.4$ contains 8 clumps, while the most magnified image shows 27 of them, including small and compact star clusters.  Thus, the physical resolution reached within each galaxy has an important effect on the characterisation of their clump and cluster population. Depending on the physical resolution reached, i.e., the lensing magnification, we detect and characterise different scales of stellar structures.

\begin{figure*}
	\includegraphics[width=18cm]{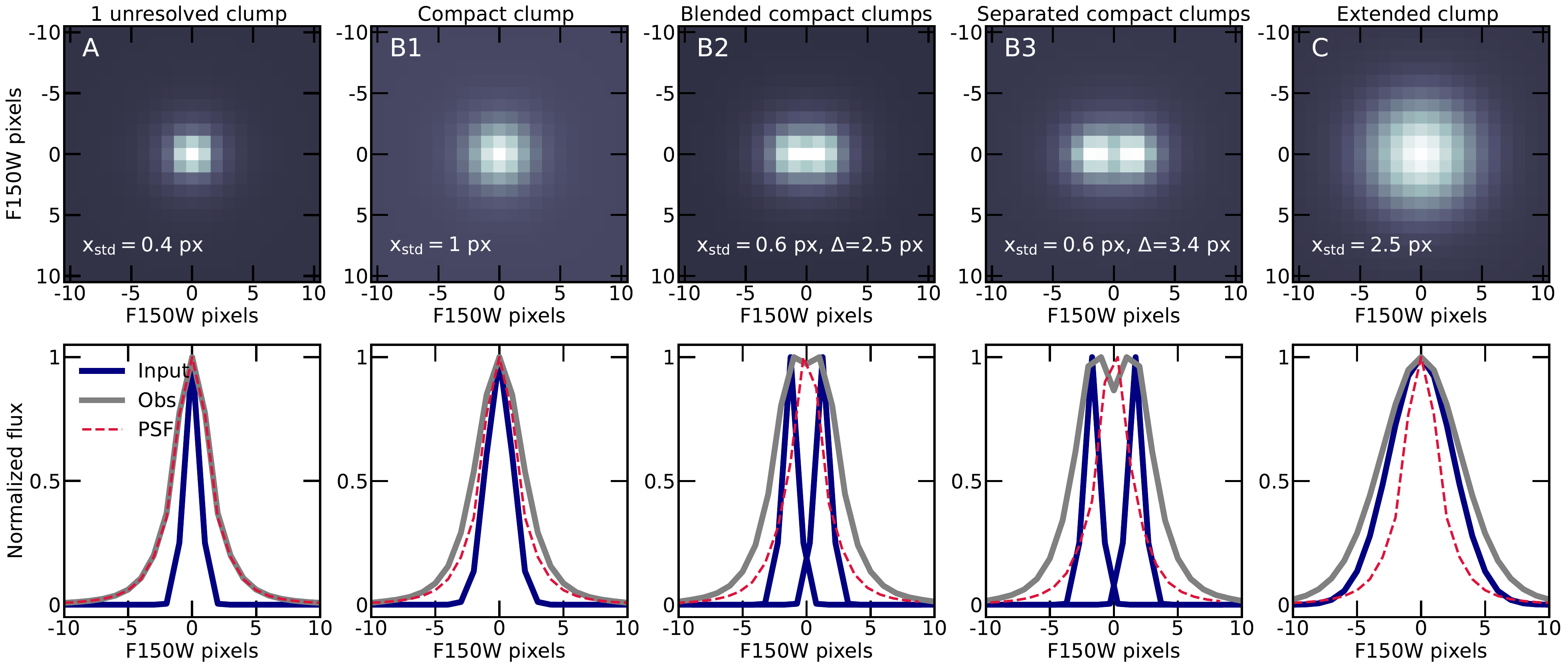}
    \caption{Illustration of the impact of resolution on the clump identification. We show 5 different cases of unresolved and resolved clumps given their intrinsic profile shape compared to the NIRCam PSF in the filter F150W. The top row presents the simulations of observed NIRCam/F150W frames with no noise. The bottom row shows the 1-D profiles of the input clump (blue), the F150W PSF (red), and the observed profile (grey). The clumps are modelled with circular 2-D Gaussian profiles. From left to right, we illustrate 5 cases in three categories: A. unresolved (228 clumps), B. compact (694 clumps) and C. extended (838 clumps). }
    \label{fig:resolution}
\end{figure*}

\noindent In our sample we do not have examples of multiple images of the same galaxies with very different magnification factors. 
However, with 1956 clumps detected among 476 individual galaxies over a broad range of redshifts and magnification factors (Fig.~\ref{fig:Sample_galaxies_clumps} and \ref{fig:Magnification}), we can study the impact that resolution has on our results. In Fig.~\ref{fig:Magnification}, we clearly see that the magnification  strongly determine our capacity to resolve very small (i.e., with $\rm R_{eff}<60 \ pc$) clumps at all redshifts. High magnifications (i.e, $\mu>80$) are necessary to detect low surface brightness clumps, at $z>2.5$. In Fig.~\ref{fig:resolution} we define five different cases  based on idealised simulations of clumps and depending on the resolution reached in the images. 

\noindent The case A represents the unresolved clumps. In this case, the input clump is too small to be resolved ($\rm x_{std}<0.4$ px) and the detected clump is a PSF-size point source and the size of the intrinsic component cannot be recovered with our PSF-photometry method. When searching among our sample, we find that 224 clumps ($\sim 12 \%$) comply with this case and we can only give a size upper-limit based on the FWHM of the stellar PSF (they are represented as black circled points in Fig.~\ref{fig:Magnification}, \ref{fig:Mag_Reff} and \ref{fig:densities}). These clumps will hereafter be referred to as "unresolved clumps".

\noindent The case B contains clumps that are broader than the stellar PSF but compact. In this case we recover physical size larger than the stellar PSF in their reference filter. Based on the assumption that they can be described as a 2-D elliptical Gaussian, this means that they will have an intrinsic Gaussian FWHM of at least $0.94$ pixels (i.e., a Gaussian sigma of $0.4$ px, see Sect.~\ref{sec:magnification_corrections}). This represents 652 clumps, corresponding to $\sim 38\%$ of the sample. These clumps are considered as resolved in the sense that we can recover their size, based on the assumption that they are composed of only one compact component. 

\noindent However, given the observed compactness of these structures, it is impossible to say if some of them are actually composed of two or more isolated clumps very close to each other (as in case B2). We will refer to them as "compact clumps" in the rest of the study. In particular, case B1 represents clumps extended enough ($\rm x_{std}>0.4$ px) to be resolved, the observed structure is larger than the PSF and thus the intrinsic size in measurable. However, two clumps can only be distinguished if they are separated by at least one PSF FWHM (corresponding to 3.4 pixels = 0.068 arcseconds in F150W). The case B2 represents two input clumps that are too close to be distinguishable in the observed frame ($\rm \Delta<FWHM_{PSF}=0.068''$). Therefore, this type of object will be detected as one elongated compact clump. Two individual clumps need to be separated from at least 3.4 pixels to be detected as two structures. Case B3 shows two compact-resolved clumps separated by 3.4 pixels and detected as two different components. These latter two cases are named "blended compact" and "separated compact" clumps. 

\noindent Finally, case C presents one extended clump, with an intrinsic 2D Gausssian FWHM larger than the FWHM of the stellar PSF (with $\rm FWHM_{clump}>FWHM_{PSF}$). If two distinct components were constituting these big clumps, the 2 components would be detected as 2 different clumps, similarly to case B3. We call these clumps "extended clumps" in the rest of the study. We measured 829 extended-resolved clumps in our sample representing $\sim 48 \%$ of the total sample at $z<5.5$. These extended-resolved clumps could of course also be composed of multiple very small components, including HII regions, star clusters and/or stellar regions, impossible to resolve individually with the current observations  and then could be coherent dense stellar regions hosting unresolved and undetected sub-components.

\begin{figure*}
	\includegraphics[width=18cm]{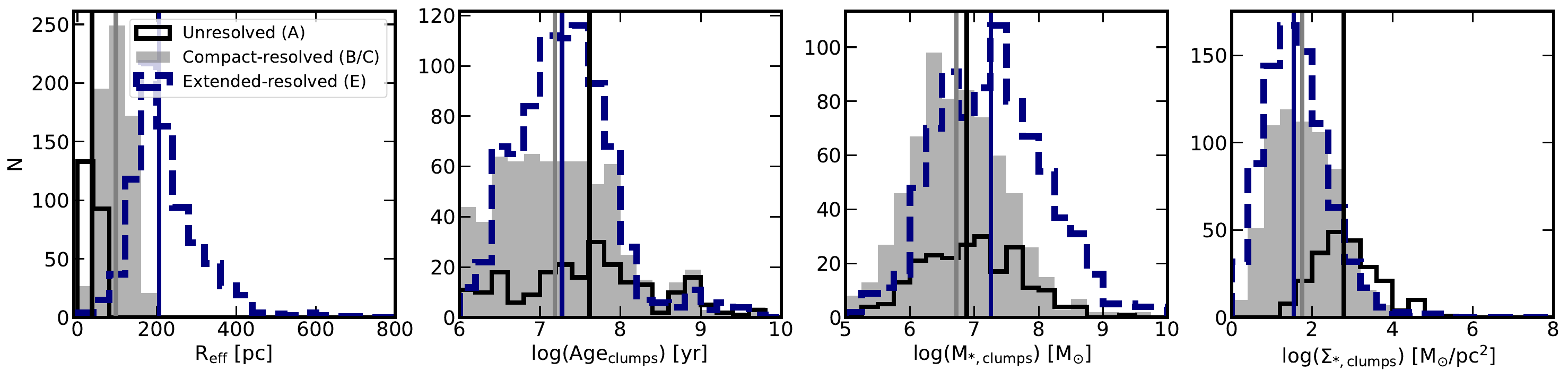}
    \caption{Distributions of clumps' effective radii, ages, stellar masses, and stellar mass surface densities in three categories of resolution (228 unresolved, 694 compact and 838 extended). We show in  black the unresolved clumps (with $\rm x_{std}<0.4$ px, corresponding to the case A in Fig.~\ref{fig:resolution}), in grey the resolved clumps with $\rm x_{std}>0.4$ px but $\rm \Delta<FWHM_{PSF}$ (corresponding to the cases B1 and B2 in Fig.~\ref{fig:resolution}), and in blue the resolved and isolated clumps (with $\rm \Delta>FWHM_{PSF}$, corresponding to the case C in Fig.~\ref{fig:resolution}).}
    \label{fig:resolution_histo}
\end{figure*}

\noindent Fig.~\ref{fig:resolution_histo} shows the clump properties of the three categories of clumps: unresolved, compact (including blended and separated compact clumps), and extended. As expected, the unresolved clumps have all sizes below 80 pc because depending on redshift and magnification 20-80 pc is the minimal resolution achievable in the redshift range considered. The compact clumps have all sizes lower than 200 pc (this upper limit is determined again by the achievable redshift and magnification combinations). However, the extended clumps present a large range of sizes between 20 and 700 pc depending on the redshift and the magnification, with a median of 206 pc, which is 1.5 times larger than the median size of all $0.7<z<5.5$ clumps of this sample. In total, we detect 448 extended regions larger than 200 pc, mainly detected at $z<4$ in galaxies with $\rm M_{*,gal}>10^8 \ M_{\odot}$. As expected, the resolution reached is also affecting the measured stellar mass surface density distributions of clumps. The unresolved clumps present very high densities (and these are the lower limits, their real densities are significantly higher), with a median value of $\rm log(\Sigma_{*,clumps}/[M_{\odot}/pc^2])=2.8$, while the median of the whole clump sample is $\rm log(\Sigma_{*,clumps}/[M_{\odot}/pc^2])=1.8$. The compact and extended clumps span the full range of measured densities, with median values of $\rm log(\Sigma_{*,clumps}/[M_{\odot}/pc^2])=1.7$ and $\rm log(\Sigma_{*,clumps}/[M_{\odot}/pc^2])=1.5$, respectively, comparable to the whole sample median.

\noindent On the other hand, the physical resolution is not affecting the mass-weighted age distribution. We recover young and old ages in the three categories of clumps, with very similar median values. We find more massive clumps in the extended-resolved category, which is expected as they correspond also to the larger clumps. Age is less affected by the resolution.

\noindent With no magnification (i.e, $\mu=1$), the minimal measurable size for a single Gaussian component in the filter F150W increases from 71 to 85 pc from $z=0.7$ to $z=1.4$ and then decrease down to 58 at $z=5.5$. In the whole sample we detect 424 clumps smaller than the minimal size achievable at the corresponding redshift without magnification. These structures, representing $25\%$ of the sample, would have been either blended with multiple other clumps or detect with larger recovered size without lensing magnification. JWST has not only enabled us to better resolve clumps in our studies. The combination of JWST sensitivity and gravitational lensing has also enabled us to detect many more clumps than seen with HST. The increase in detection can be quantified in our sample. We used our sample of JWST extracted clumps to verify the detection of the clump light in the HST imaging data. We consider the clump as detected if it was recovered with a $\rm S/N>2$ in three HST bands and confirmed by visual inspection.  In total only 30\% of our clumps have been detected in in the existing HST images. Moreover, with only 3 HST broadband it would have bee impossible to derive any other physical property.

\subsection{What type of clumps do we observe?}
\label{sec:type_clumps}

Throughout this study, we have defined a clump as an observational entity whose physical nature depends on the physical resolution reached in the observations. Depending on the distance and magnification, one probes different scales of stellar structures: star clusters, small and large star-forming regions, AGNs, bulges, galaxies. 

\begin{figure*}
	\includegraphics[width=18cm]{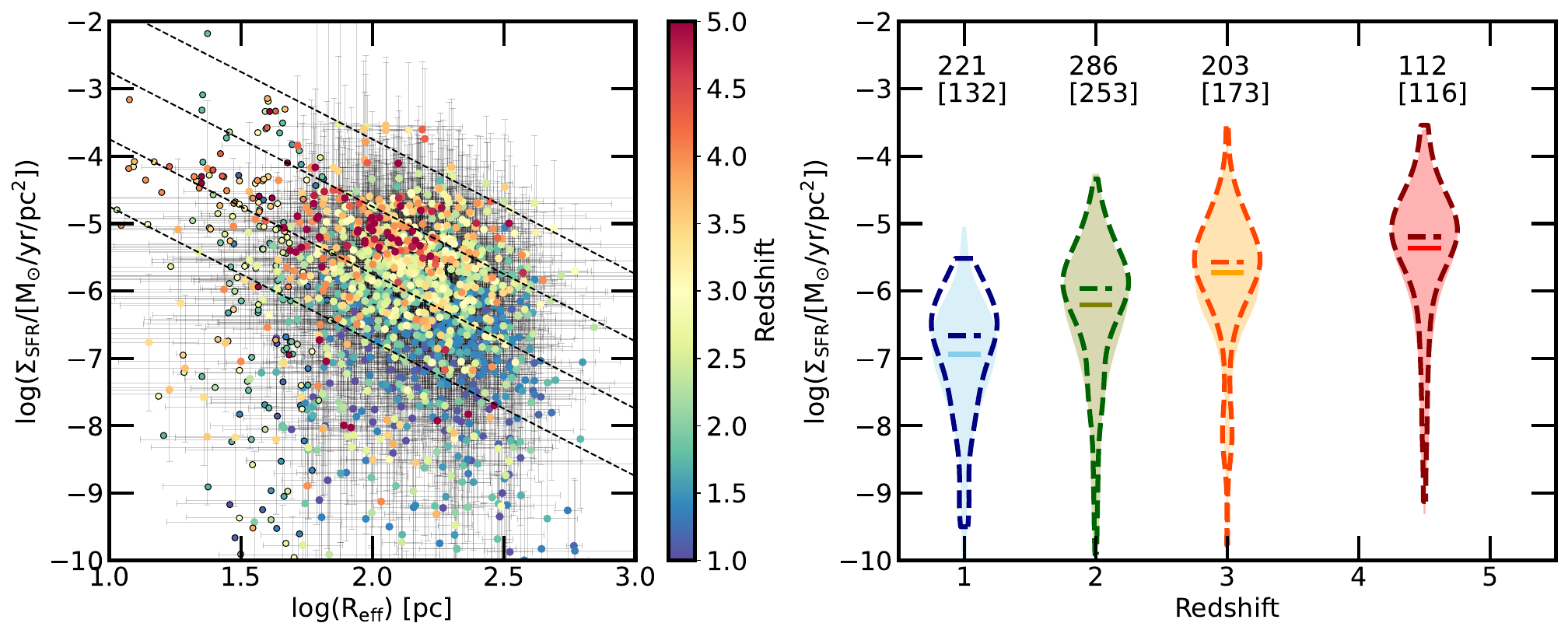}
    \caption{Left: Clump SFR surface density as a function of their effective radius colour-coded with redshift. The dashed lines represent SFR isochrones at 10, 1, 0.1, and 0.01 $\rm M_{\odot}/yr$. The points circled in black are upper limit in size. Right: Violin distribution of clump SFR surface density of clumps divided in the four redshift bins. The filled violins represent  the extended-resolved clumps in each redshift bin, while the dashed violins contain the compact-resolved clumps. The numbers of compact-resolved clumps within each bins are indicated in black and the numbers of extended-resolved clumps are indicated between brackets.
    }
    \label{fig:densities}
\end{figure*}

\noindent We present the properties of 1751 individual stellar clumps at $0.7<z<5.5$. These clumps range from 1 to 700 pc in effective radius and reach stellar mass surface densities from $\rm log(\Sigma_{*,clumps}/[M_{\odot}/pc^2])=0$ to $\rm log(\Sigma_{*,clumps}/[M_{\odot}/pc^2])=5.2 $ at all redshifts (Fig.~\ref{fig:densities}). Among these clumps, 88\% of them are resolved (cf Fig.~\ref{fig:resolution} and ~\ref{fig:resolution_histo}), and 48\% are extended-resolved. The clump properties from our sample are similar to the SMACS0723 stellar clumps presented in \citetalias{Claeyssens2023}.

\noindent We detect only 34 very compact clumps with a size smaller than 20 pc, i.e. at the scales of star clusters; with all of them being unresolved (i.e., their size is an upper-limit). These very compact clumps are detected at all redshift and have stellar masses ranging from $\rm 10^{4.5}$ to $\rm 10^{8.5} \ M_{\odot}$. They have mass-weighted ages ranging from 1 Myr to 1 Gyr, and SFR going from $10^{-4}$ to 10 $\rm M_{\odot}~yr^{-1}$. They present on average higher stellar surface densities and they will be intrinsically higher, since these densities are lower limits (similarly observed in \citealt{Messa2024}). These clumps could be consistent with star cluster candidates at high redshift.

\noindent The majority of clumps we detect are large stellar regions with 1100 clumps with $\rm R_{eff}\geq100$ pc and among those, almost all of them are extended-resolved, which we assume to be be consistent with coherent stellar structures, probably hosting star clusters within them. We detect more of these large structures in the low-redshift galaxies (cf Fig.~\ref{fig:clumps_redshift}).

\noindent  As shown in Fig.~\ref{fig:host_galaxies}, galaxy that are currently forming a significant fraction of their mass with log(sSFR/[$\rm yr^{-1}$])$>-8.5$, have a larger fraction of their mass residing within clumps (typically 45\%) suggesting that stellar clumps are a major mode for the build up of the stellar mass of there early galaxies. This is particularly true at $z>2$. Given their high densities and stellar mass, most of the clumps should host star clusters. In the local universe, the average stellar surface density of bound star clusters detected is $10^{2.4} \ \rm M_{\odot}/pc^2$ (\citealt{BG2021}). We measure a significant fraction of clumps with stellar mass surface densities larger than this value (13\% at $0.7<z<1.5$, 21\% at $1.5<z<2.5$, 29\% at $2.5<z<3.5$, and 41\% at $3.5<z<5.5$). The very high stellar densities measured could indicate that star clusters are forming in more extreme environments at these redshift.

\noindent Even if we do not identify them in the current analysis, we cannot exclude a potential contamination of the sample by AGNs. Indeed, several AGNs have been detected with JWST with clump-like properties in size and density, especially at very high redshifts ($z>6$), including one in the field of A2744 that we have removed from our sample \citep{Furtak2024_AGN}. In total, 260 AGN candidates at $4<z<9$ have been identified in the JWST/NIRCam images of A2744 (\citealt{Kokorev2024}).

\begin{figure}
	\includegraphics[width=9cm]{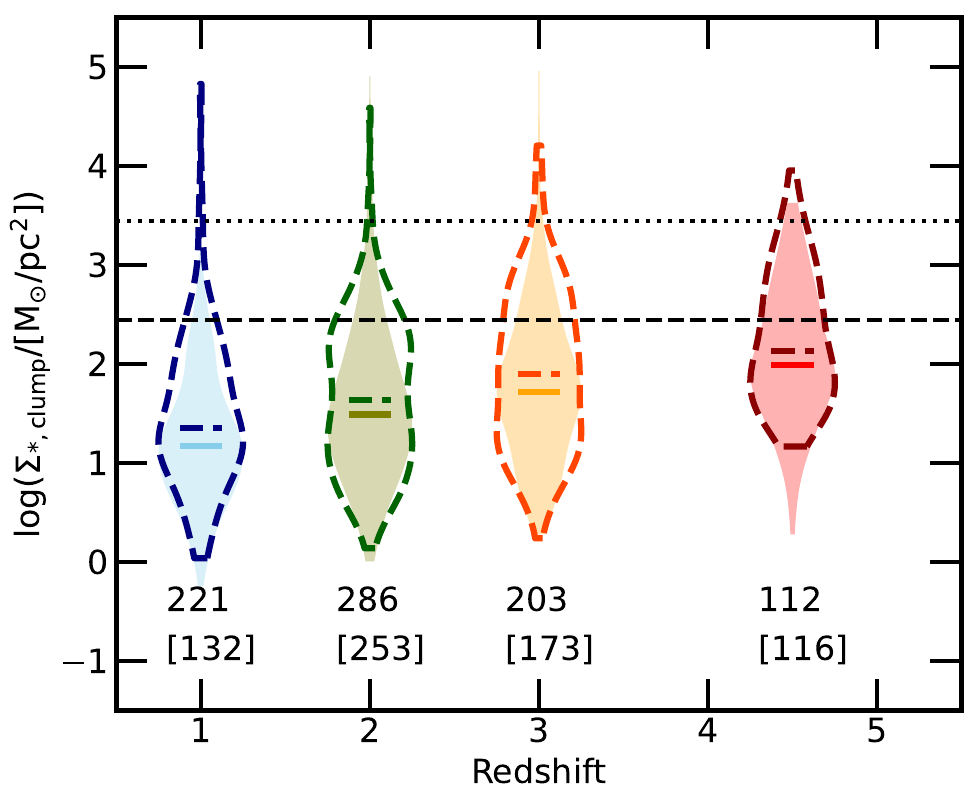}
    \caption{Violin distribution of clump stellar mass surface density of clumps divided in the four redshift bins. The filled violins represent the clumps extended-resolved in each redshift bin, while the dashed violins contain the compact-extended clumps (see Sect.~\ref{sec:resolution} for the definition of these two categories). 
    The dashed black horizontal line represents the average surface density of a nearby massive cluster, from \citet{BG2021}. The dotted black line represents the typical surface density of a globular cluster with $\rm R_{eff} = 3 \ pc$ and $\rm M_* = 10^{5.2} \ M_{\odot}$. The numbers of compact-resolved clumps within each bins are indicated in black and the numbers of extended-resolved clumps are indicated between brackets.
    }
    \label{fig:densities_mass}
\end{figure}

\subsection{Do clump properties evolve with redshift?}
\label{sec:redshift_evolution}

Studies of stellar clumps in HST and JWST observations have tried to measure the redshift evolution of clump properties. \citet{Livermore2015} reported that the median SFR and SFR surface density of clumps increased with redshift. \citet{Messa2024}, combining different samples of HST and JWST highly magnified clumps from the literature, have measured an increase of the median clump SFR surface density with redshift from $z=0$ to $z=6$. They measured a strong evolution between the local clumps at $z=0$ and the clumps at $3.5<z<5.5$. The evolution of stellar mass surface densities with redshift is more subtle, with clumps reaching high stellar densities more easily at higher redshift.

\noindent In our sample, probing a wider range of magnification, we observe a very weak evolution of the stellar surface densities from $z=0.7$ to $z=5.5$ with a median of $\rm log(\Sigma_{*,clumps} /[M_{\odot}/pc^2]) =1.3^{+0.7}_{-0.37}$ at $0.7<z<1.5$ against  $\rm log(\Sigma_{*,clumps} /[M_{\odot}/pc^2]) =2.2^{+0.6}_{-0.4}$ at $3.5<z<5.5$, while we measured individual clump densities ranging from 0 to 5 in every redshift bin. As the total clumps sample is mixing unresolved, compact and extended clumps in every redshift bins, we compute the stellar surface density distributions with only the resolved clumps, divided in two categories: compact-resolved and extended-resolved (cf Sect.~\ref{sec:resolution}) presented in Fig~\ref{fig:densities_mass}. The stellar surface density redshift evolution is very similar in the two subsamples with the compact-resolved clumps having slightly higher densities. Overall, it is difficult to disentangle the effect of completeness (which would mainly affects the detection of high-redshift low surface density clumps), and broad size ranges as a function of redshift. In spite of the shallow increase, it remains remarkable that the clumps on 10s parsec scales can reach density of star clusters and globular clusters observed in the local universe, telling us that star clusters within them are forming in very high stellar density regions. Indeed, when resolving star cluster physical scales, extremely dense star clusters are detected in very high-redshift galaxies with JWST, e.g. in the Cosmic gems at $z=10$ (with five young $\rm R_{eff}<2 \ pc$ star clusters with  $\rm \Sigma_{*}>10^5 \ M_{\odot}/pc^2$, \citealp{Adamo2024} ), in the Firefly Sparkler at $\rm z=8.3$ \citep{mowla2024}, in the Cosmic Archipelago at $\rm z=6.1$ \citep{messa2024_D1T1} and in the Sunrise arc at $z=6$ \citep{Vanzella2023}, all hosting several clusters with $\rm R_{eff}<10~pc$ and reaching densities $\rm \Sigma_{*}>10^3~M_{\odot}/pc^2$.

\noindent Fig~\ref{fig:densities} shows the clumps SFR surface densities ($\rm \Sigma_{SFR}$) as a function of the $\rm R_{eff}$ color-coded with redshift in the left panel, and the $\rm \Sigma_{SFR}$ violin distributions in each redshift bins considering only the spatially resolved clumps. We observe a strong redshift evolution of the $\rm \Sigma_{SFR}$ distributions, $3.5<z<5.5$ clumps are concentrated at high SFR densities with 80\% of the clumps having $\rm log(\Sigma_{SFR}/[M_{\odot}/yr/pc^2])>-6$ against only 10\% at $0.7<z<1.5$. The median $\rm \Sigma_{SFR}$ evolves from $\rm log(\Sigma_{SFR}/[M_{\odot}/yr/pc^2])= \ -6.8^{+0.4}_{-0.7}$ at $0.7<z<1.5$ to $\rm log(\Sigma_{SFR}/[M_{\odot}/yr/pc^2])= \ -5.2^{+0.5}_{-0.5}$ at $3.5<z<5.5$. This evolution is comparable to the findings from \citet{Messa2024}. The redshift evolution of the clump SFR density is similar for compact-resolved and extended-resolved clumps, with the compact-resolved clumps having slightly higher SFR densities.

\noindent As we already discussed in Sect.~\ref{sec:resolution} and Sect.~\ref{sec:magnification}, the redshift evolution of the clumps densities, especially the stellar and SFR densities, can be strongly affected by the completeness issue and/or the surface brightness limit at higher redshift. We find very dense clumps within the sample, reaching stellar mass surface densities of $\rm \Sigma_{*} \sim 10^4 \ \rm M_{\odot}/pc^2$  and SFR surface densities of $\rm \Sigma_{SFR} \sim 10^{-4} \ \rm M_{\odot}/yr/pc^2$ at $z>1.5$. 

\noindent The SFR surface density evolution can also be explained if we consider, as already seen in Fig.~\ref{fig:clumps_redshift} and ~\ref{fig:migration}, that clump are becoming increasingly younger at higher redshift and more SFR is taking place in high redshift clumps during the last 10 Myr, thus making the redshift evolution of SFR density stronger. The stellar mass density evolution is showing that very dense clumps become more common as we move to higher redshift. Higher stellar densities will have implications for survival timescale of the clumps as denser clumps should survive longer as predicted in simulations (\citealt{Ceverino2023}). 

\noindent In Fig.~\ref{fig:clumps_redshift}, we present the clump properties in four redshift bins. Even if we detect a wide range of clump parameters at all redshifts, we observe some significant redshift evolution. We measured significant average variations of the SFR, size, and age across cosmic time. Clumps at $3.5<z<5.5$ are, on average, younger, more star-forming, more compact, and have a lower metallicity, while clumps within $0.7<z<3.5$ galaxies can reach older ages (with some clumps begin $>300 \rm \ Myr$) and larger sizes.

\noindent As shown in Fig.~\ref{fig:host_galaxies}, the clump properties are also linked to their host galaxy properties and then the redshift evolution of clump properties follows the general trend of galaxy evolution through cosmic time; with time galaxies are growing in mass and size, forming more metals and more morphological structures (such as proto-bulges and bulges, as well as spiral arms and bars). \citet{Renaud2024} have also shown that gas-rich galaxies (with a gas fraction of $\rm f_{gas}=40\%$) will host on average more massive and more dense clumps and allow for the formation of extreme clumps not found at low-gas fraction galaxies ($\rm f_{gas}=10\%$).

\subsection{How do clumps form?}
\label{sec:clumps_formation}

As already discussed in the introduction different scenarios are considered to explain clump formation. In the {\it in situ} scenario, the clumps are formed by the large-scale fragmentation of gaseous disks due to violent disk instabilities caused by intense cold gas accretion flows. This scenario has been supported by several numerical simulations (\citealt{Ceverino2012, Bournaud2014, Tamburello2015, Behrendt2016, Madelker2017}).  Moreover, observational works have shown that the galaxies across similar redshift range as in our sample, covering the peak of cosmic star formation ($1<z<5.5$), are characterised by a more turbulent and denser ISM and with stronger pressure (\citealt{Wisnioski2015}) and higher molecular gas mass fractions (\citealt{Tacconi2018,Liu2019,DZ2020}). 
If the clump formation is driven by fragmentation, the distribution of clumps is expected to follow a hierarchical structure as gravitational fragmentation and turbulence compression are scale-free processes, then both gas and stars are expected to follow continuous density distributions that are described by log-normal functions (\citealt{Elmegreen2005}). 

\begin{figure*}
	\includegraphics[width=18cm]{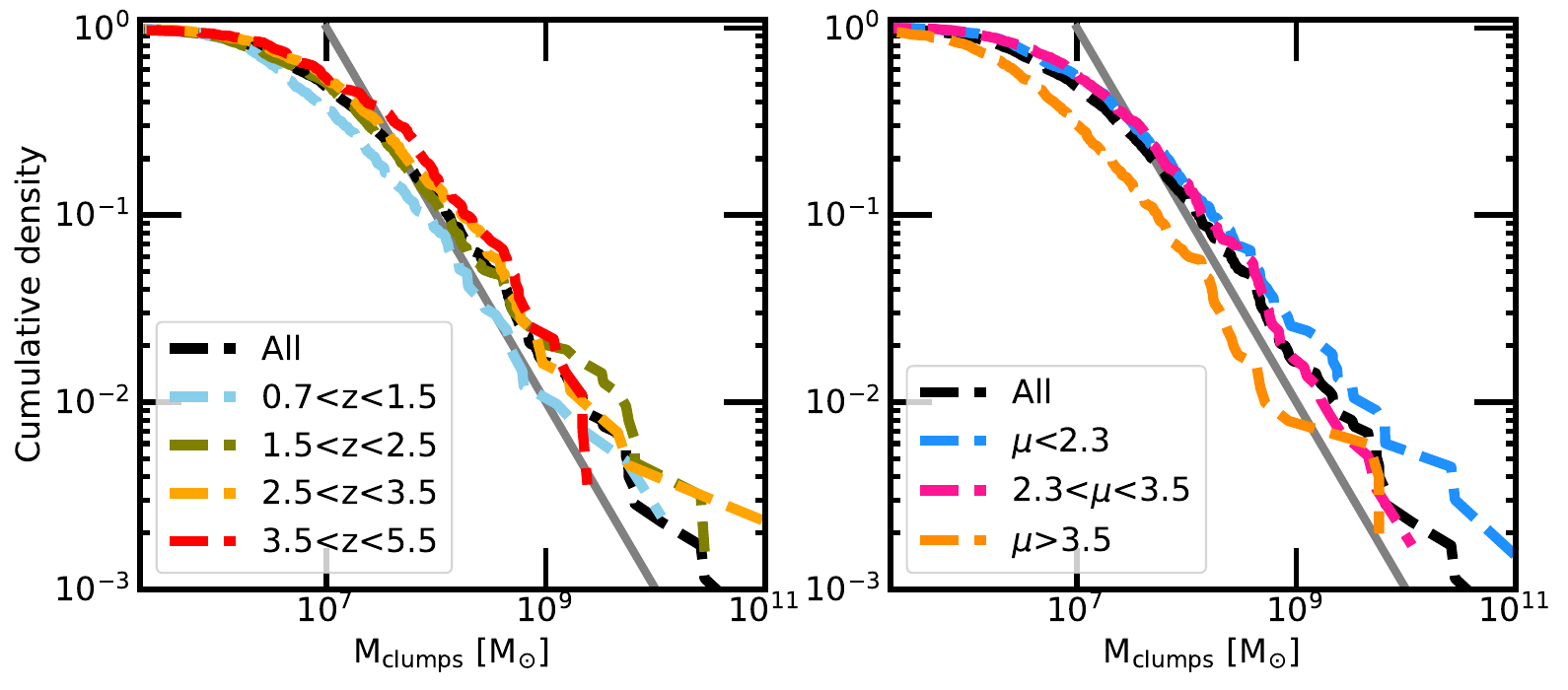}
    \caption{From left to right: Cumulative stellar mass distributions of clumps divided in four redshift bins and three magnification bins. In each panel, the black dashed distribution corresponds to the full clump sample, whereas the coloured distributions represent the different bins. The grey solid line represents a the shape of a cumulative distribution corresponding to a power-law distribution with a slope $\alpha=-2$. }
    \label{fig:cumulative_distrib}
\end{figure*}

\noindent To investigate the formation mechanism of clumps, we computed the cumulative mass distributions of clumps (see Fig.~\ref{fig:cumulative_distrib}). In the literature there is a consensus that both the initial star cluster mass and luminosity functions in local galaxies are well-described by a power-law slope of approximately $-2$ (e.g., \citealt{Whitmore1999, Larsen2002, Bik2003, Gieles2006, Chandar2014,Whitmore2014,Adamo2017,Messa2018,Mok2019,Mok2020}). In local galaxies, gas fragmentation is driven by turbulence and results in star-forming regions (star clusters complexes) that are hierarchically organised from pc to kpc scales that follow a stellar mass distribution close to a power-law of slope $\alpha\sim-2.0$. 
Until now, the stellar mass function of star-forming clumps at $z>1$ has remained poorly constrained (\citealt{DZ2018}), mainly because of the poor statistics available for such objects. Indeed, the major challenge for this type of clump analysis resides in compiling a well-controlled sample of high-redshift clumps with a well-understood effect of the spatial resolution, with stellar masses derived in a homogeneous way, and, most importantly, with a full understanding of the sensitivity effect on the clump detection. Moreover, the determination of the clump mass function has suffered from the low number statistics, between a few up to tens of clumps per galaxy, even with the help of lens magnification.

\noindent In this work, we have collected the largest sample of high-redshift clumps ($1751$ clumps) covering the peak of the cosmic star formation ($0.7<z<5.5$). Thanks to the combination of the lens magnification and high sensitivity of the JWST/NIRCam instrument we pushed the mass detection limit of high-redshift clumps down to $10^5 \ \rm M_{\odot}$. We computed the cumulative mass distributions of the clumps in different bins of redshift and magnification (see Fig.~\ref{fig:cumulative_distrib}). We notice a very stable distribution with redshift, consistent with a power-law distribution with a slope of $-2$ (shown in grey in the figure). The incompleteness of our sample at low surface brightness clumps (i.e., low mass) produces the shallow slope at low masses observed in Fig.~\ref{fig:cumulative_distrib}. In general, we conclude that the consistency with a $-2$ power-law supports the in-situ scenario for clump formation at all redshifts between $z=0.7$ and $z=5.5$.
\noindent Therefore, there may be a continuity in the way that the largest coherent star-forming units are formed in galaxies at $z \sim 5.5$ down to $z=0$. Interestingly, we notice that the upper mass-end of the distribution evolves with redshift, with the most massive clumps seen in the redshift bins sampling the cosmic noon (cf Sect.~\ref{sec:clumps_redshift}).  

\noindent When dividing the sample according to the clump magnification, we notice a slight evolution of the cumulative mass distributions. The most massive clumps (with $\rm M_{clumps}>10^9 \ M_{\odot}$) are only detected at $\mu<2.3$. This can be easily explained by resolution effect, as discussed in Sect.~\ref{sec:resolution}, however, we see that the global distribution is still consistent with a $-2$ power-law.

\noindent This result is also in agreement with the increasing number of disk-dominated galaxies with redshift discovered by JWST (\citealt{Ferreira2022,Ferreira2023,Kartaltepe2023}).  While the presence of disks, and the consistency with $-2$ power-law mass distributions points toward the {\it in situ} scenario, as we move to higher redshift, we cannot however exclude that increasing number of minor merger events is taking place. Further studies and larger statistics are necessary to probe the dominant formation mechanism as a function of redshift.

\subsection{Clump survival timescale and migration}
\label{sec:migration}

In Fig.~\ref{fig:clumps_redshift}, we see that most of the clumps are younger than 300 Myr, which means that they have been formed in less than 300 Myr, i.e. no later than $z\sim4$ for clumps detected at $z<3.5$. This implies that most of the clumps observed at $z>3.5$ would not survive long enough to be observed at $z<3.5$ and that the majority cosmic noon clumps observed have been formed during the cosmic noon period. The mass-weighted stellar ages measured in our sample are very consistent with the recent simulations which find typical clump ages of a few hundreds Myr (\citealt{Dekel2021, Ceverino2023}). In these simulations, the clumps are shortly disrupted by strong feedback. However, the long-lived clumps (as defined as older than 50-100 Myr in their work) can migrate within the disk, drive gas inflows towards the galaxy center, and contribute to the bulge formation. Only the densest clumps are able to survive feedback and become long-lived clumps. Moreover, this migration effect would shorten their survival  timescale and they would disappear in a few hundreds Myr, which corresponds to the time they would need to spiral into the center of galaxies (\citealt{Dekel2021,Dekel2022,Ceverino2023}). \\

\noindent In Fig.~\ref{fig:migration}, we present the clump properties distributions as a function of their location within their host galaxy. If the clumps form mostly in the outer part of the disk and then all migrate, this effect would produce a measurable age gradient within the galaxies with the most outskirt clumps being younger than clumps closer to the centers of galaxies. However, on average, in the whole clump sample, we do not observe such a galactocentric distance trend with the clump age, at any redshift and any galactic stellar mass. Interestingly, we measured a small but significant fraction of clumps (10\%, see Fig.~\ref{fig:clumps_redshift}) with mass-weighted ages higher than the typical time for migration quoted in the simulation works (between 300 and 500 Myr). Several of these clumps are located in the center of galaxies, and present mass and size consistent with being (proto-)bulges of these galaxies. However, we also observe such old clumps at larger distances (i.e., at 1 kpc or more from the center of their host galaxy). These results suggest that a small fraction of clumps could survive longer than expected even far from the center of the galaxy. 

\noindent Using the detection of 3187 clumps (from 1269 galaxies from the HST CANDELS-GOODS fields) at $z=0.5-3$, \cite{Guo2018} and \cite{Dekel2021} have measured age-galactocentric distance gradients similar to the one determined in the simulations. A few other studies have examined radial gradients of clump properties with galactocentric distance in small samples of clumps and galaxies (\citealt{FS2011, Guo2012, Soto2017, Zanella2019}). They all find evidence for older, redder, more massive clumps closer to the disc center, in qualitative agreement with the findings of \citep{Guo2018} and \citep{Dekel2021}, and the predicted migration-induced radial gradients of clumps with age$>50-100$ Myr. However, all these samples study kpc-size clumps detected in non-lensed observations of galaxies. It is difficult to estimate how the blending of clumps in non-lensed galaxies observations could affect the age measurement, and then the measured gradients, in a systematic way. No significant migration signs have been evidenced in lensed clump samples. \\

\noindent In contrast to the ages, we measured a very significant evolution of the clump stellar mass fraction towards the center of galaxies at every redshift and galactic mass (Fig.~\ref{fig:migration}). The same trend has been observed in most of the clump samples (\citealt{Guo2012, Shibuya2016,Soto2017,Guo2018,Dekel2021,Kalita2024}). Multiple effects could participate in producing this trend. More massive clumps towards the centers of galaxies could simply be due to the formation of more massive clumps in the centers of  disks. Alternatively, clumps could grow in mass via the merging of smaller clumps into more massive structures while migrating towards the centers of galaxies. Moreover, combined with the migration, many simulations (\citealt{Dekel2021, Ceverino2023}) suggest that only the most massive and dense clumps will survive disruption due to stellar feedback. This will lead to more massive clumps being closer to the centers. On the other hand, the mass gradient may simply be an effect of higher concentration of gas close to the galaxy core (e.g., \citealt{Elbaz2018, Puglisi2021, Gomez-Guijarro2022}) and the clumps actually would not survive long enough to migrate across the galactic disk. \\
Then, the mass-distance gradient observed in Fig.~\ref{fig:migration} would be due to variation of the SFR across the galaxy from the inner to the outer regions. \\

\begin{figure*}
	\includegraphics[width=18cm]{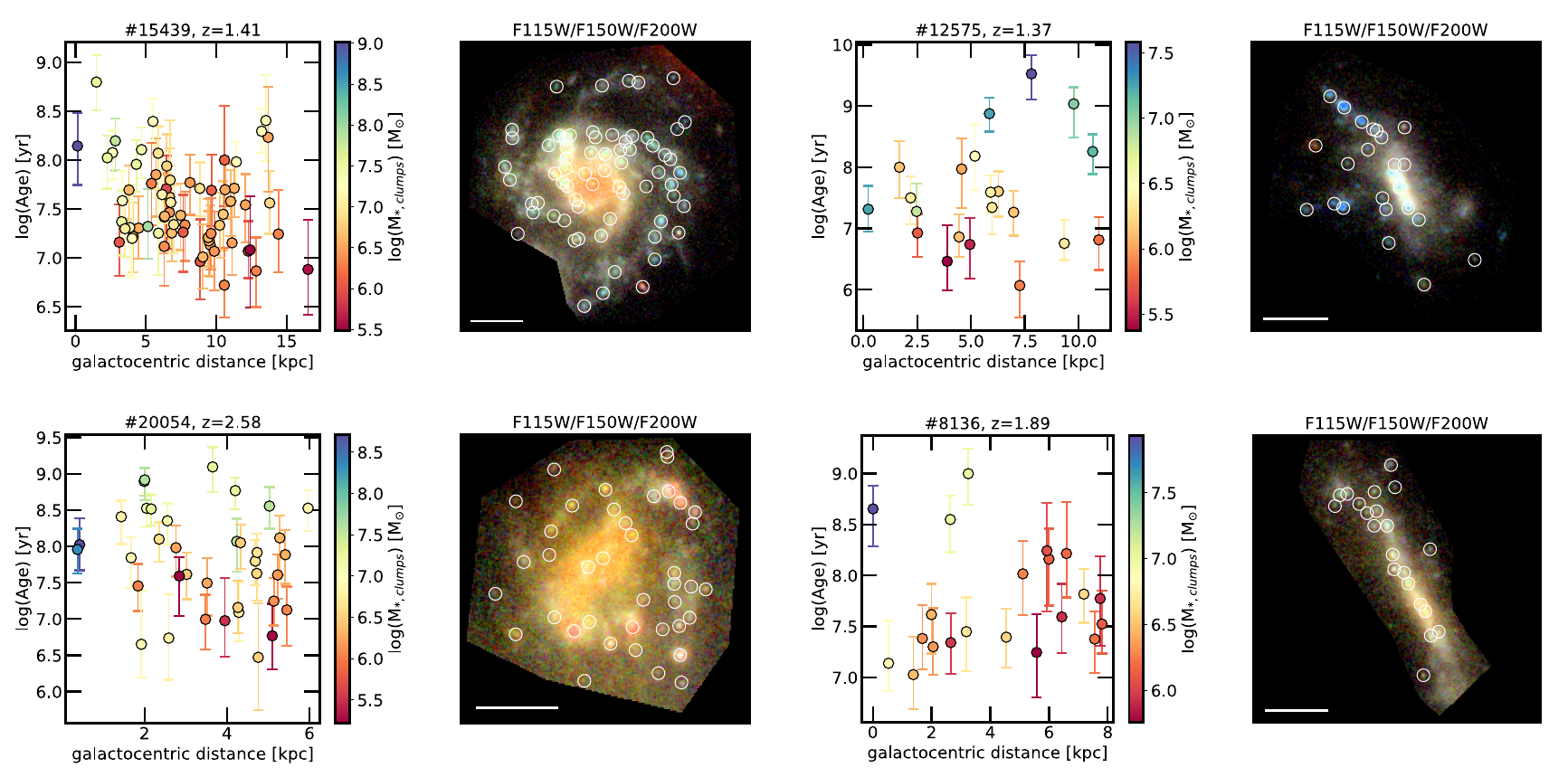}
    \caption{Clump mass-weighted ages as a function of the galactocentric distance for four galaxies with more than 15 clumps. The points are colour-coded by stellar mass. The ID and redshift of each galaxy are indicated on top of the plot. We show also a RGB image (with R=F200W, G=F150W and B=F115W) of the galaxies with the clump locations indicated by white circle. The white scales indicate 0.2 arcseconds. Galaxies on the left are face-one spiral galaxies at $z=1.41$ and $z=2.58$, galaxies on the right are two very irregular galaxies, potentially edge-on.}
    \label{fig:migration_indiv}
\end{figure*}

\noindent In Fig.~\ref{fig:migration_indiv}, we present the same figures for 4 individual galaxies. We selected these galaxies among the most clumpy galaxies from the sample to illustrate the impact of projection effects on the determination of migration trends. All these galaxies are at $z<2.5$. We do not measure any significant trend in age in any of these galaxies. However, analysing the galactocentric distances of individual galaxies, the inclination bias associated with the 2D projections of the clumps within the galaxy might strongly affect the age measurements for some of our clumps by contaminating the total flux of the clumps by light belonging to independent structures located on the same line of sight (such as the galactic dust and gas emissions, star clusters, etc). Indeed, we present 2 face-on spiral galaxies in Fig.~\ref{fig:migration_indiv} (\#15439 and \#20054) which do not show any obvious sign of on-going merger and/or high quantity of dust. However, we measure only a very weak age-distance dependence within these galaxies. Each of them contain both young ($<20$ Myr) and old ($>100$ Myr) clumps with a slightly larger fraction of the last ones close to the center of the galaxies. The older clumps close to the center are also the more massive ones, on average. Moreover, their central clumps are older than the median age within the disk and significantly more massive, which indicates that they are consistent with being galactic bulges. Therefore, the age and mass distributions of the clumps within these galaxies show a weak sign of migration. The detection of young clumps distributed everywhere within the galaxies tends to indicate that clumps and star formation is still ongoing within. The two other galaxies shown (\#12575 and \#8136) are edge-on galaxies, with no clear mass and/or age trends detected. 
In addition to the orientation of the galaxies, major mergers will completely shuffle the spatial distribution of the clumps and could erase a potential age gradient already in place before the merger event. Finally, we observe also among our sample very red galaxies suggesting a significant amount of dust. It is well-known that the presence of dust can significantly affect the age measurements due to the strong dust-age degeneracy observed when fitting galaxy/clumps SED. To summarise, multiple observational or physical biases can affect the age measurements of some clumps within galaxies and will make the age-distance diagrams difficult to interpret.

\section{Conclusions}
\label{sec:conclusion}

In this study, we have analysed 1956 individual clumps extracted from 476 individual galaxies at $0.7<z<10$ using the recently acquired JWST/NIRCam data of the galaxy cluster Abell 2744. The majority of the clumps ($\sim 70\%$) studied in this work has gone so far undetected from the deep HST data available for this cluster. 

\noindent Galaxies have been selected according to their number of clumps and the availability of a robust redshift estimation, either from photometric or spectroscopic observations. The detection of clumps has been done in two steps. The initial extraction has been performed with the software SExtractor and then a visual inspection on multiple JWST/NIRCam images of the selected galaxies has been performed to include fainter clumps missed and remove false detections. Clumps sizes have been measured on a reference filter depending on the redshift of the galaxies and corrected from the lensing magnification. Clumps photometry has been extracted from 20 JWST/NIRCam filters. Clumps properties such as mass-weighted ages, SFR, stellar mass, dust extinction, stellar surface densities and metallicity have been derived from SED fitting using the code Bagpipes with short SFH. The combination 20 JWST/NIRCam broad and medium-band filters provide robust estimations of these parameters. \\

\noindent The reionisation era clumps at $5.5 \leq z<10$ will be presented in a separate work. We restricted the clumps physical properties study to the objects with $z<5.5$. From the analysis of these 1751 clumps we showed that:  \\

\begin{itemize}

\item  Thanks to the lensing magnifications, we break down the star-forming regions down to star clusters scales in size and mass. Among our sample we detect different types of clumps. With effective radius ranging from 1 to 700 pc, we probe, depending on the distance and magnification, individual star clusters candidates, compact star-forming regions and diffuse star forming regions. \\

\item We classified our clumps into three categories: unresolved (12\%), compact (38\%) and extended (48\%) depending on the physical resolution reached and the lensing magnification. The late one impacts strongly the recovered size, brightness and stellar mass of the clumps, while the densities are only very weakly affected. Considering only resolved clumps, the physical resolution reached does not impact the ages and densities distributions. \\

\item The clumps parameters evolve with redshift: from $z=0.7$ to $z=5.5$  clumps are on average smaller, denser (both in stellar surface density and SFR density) and younger. However, we measure similar stellar mass and dust extinction distributions in the four redshift bins. We measure a clear SFR and densities evolution from $z=0.7$ to $z=5.5$ considering only resolved clumps. The very high densities of high-redshift clumps suggests very dense clumps become more common at $3.5 \leq z<5.5$ and thus that star clusters are forming in environments that are more extreme as we move to higher redshift. The SFR density evolution shows that more SFR is taking place in the high redshift clumps in the last 10 Myr.  \\

\item The clumps properties are strongly linked to their host galaxy properties. Massive galaxies are hosting more massive clumps and on average these clumps contain older stellar population. The most massive galaxies are also hosting more massive central clumps which are (proto-)bulge candidates for the most evolved galaxies. Moreover, the fraction of mass from the galaxy contained within the clumps  ($\rm F_{Mass}$) is directly linked to the total galactic stellar mass and the specific SFR of the whole galaxy. More active ($\rm sSFR>10^{-8.5}$), low-mass ( $\rm M_{*,gal}<10^{8.5} \ \rm M_{\odot}$) galaxies present on average a higher $\rm F_{Mass}$. This trend is directly linked to the redshift evolution of galaxies, as we find more low-mass and very active galaxies at $z>2.5$. This shows that most of the star formation activity occurs within the clumps.   \\

\item We do not measure any significant clumps age-galactocentric distance trends at any redshift. However, we measure a strong mass-galactocentric distance at all redshift. The clumps get more massive towards the galactic center. This trend supports the scenario of clumps formation within the disk and then clumps migration towards the center. This measurement is consistent with similar trends measured from other samples. This could indicate also that the clumps participate to the formation of the galactic bulges. The young ages of the large majority of the clumps (90\%
of the clumps have mass-weighted ages $<300 \rm \ Myr$) tend to support the quick disruption of the clumps either from the migration or due to strong stellar feedback as proposed also by multiple simulations. More massive clumps close to the galactic center could also indicate clumps merger during the migration and/or the formation of more massive clumps close to the center thanks to a denser gas. Analysis of individual galaxies have also showed that the orientation of the targets, the presence of dust and/or the merger history of the galaxies can affect the measured of such radial trends of clumps properties.  \\

\item  At each redshift, the cumulative stellar mass distributions of the clumps are consistent with a power-law distribution with a slope of $-2$. This support the scenario of clumps formation from fragmentation within the disk due to violent disk instabilities as proposed by the majority of the simulations. This seems to be major mode of clumps formation from $z=5.5$ to $z=0.7$. 

\end{itemize}

\noindent To conclude, deep JWST/NIRCam observations of a massive galaxy cluster region in 20 different filters, have shown the incredible leap forward that we are taking in studying star formation and breaking down star-forming structures within rapidly evolving galaxies. The combination of the unprecedented sensitivity and high resolution power of JWST with the lensing magnification is a key factor in the detection and analysis of compact stellar structures within these young galaxies, previously only detected at the FUV restframe at lower physical resolution. Our analyses and the parameters distributions can be directly confronted to the most recent numerical simulations producing clumpy galaxies. \\

\section*{Acknowledgements}
The authors thank the International Space Science Institute for sponsoring the ISSI team: "Star Formation within rapidly evolving galaxies" where many ideas discussed in this article have been brainstormed. AA and AC acknowledge support by the Swedish research council Vetenskapsradet (2021-05559). M.M. acknowledges the financial support through grant PRIN-MIUR 2020SKSTHZ. JM and IK acknowledge support by the European Union (ERC, AGENTS, 101076224). Views and opinions expressed are however those of the author(s) only and do not necessarily reflect those of the European Union or the European Research Council. Neither the European Union nor the granting authority can be held responsible for them. RPN acknowledges funding from {\it JWST} program GO-3516. Support for this work was provided by NASA through the NASA Hubble Fellowship grant HST-HF2-51515.001-A awarded by the Space Telescope Science Institute, which is operated by the Association of Universities for Research in Astronomy, Incorporated, under NASA contract NAS5-26555.

\section*{Data Availability}

All the data used in this paper can be extracted as described in section 3. The HST and JWST data products presented herein were retrieved from the Dawn JWST Archive (DJA) (\href{https://dawn-cph.github.io/dja/index.html}). DJA is an initiative of the Cosmic Dawn Center (DAWN), which is funded by the Danish National Research Foundation under grant DNRF140.



\bibliographystyle{mnras}
\bibliography{A2744_clumps} 



\appendix

\section{JWST/NIRCam medium-band observations}
\label{sec:appendix}

\begin{figure*}
	\includegraphics[width=18cm]{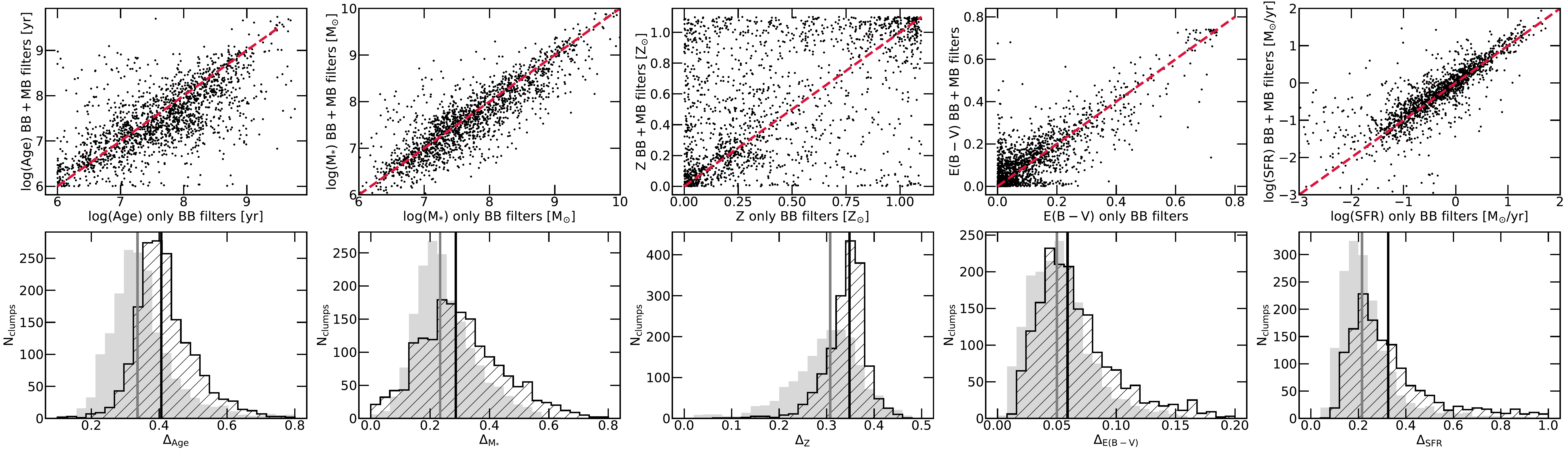}
    \caption{Comparison of the clumps parameters obtained from SED fitting based on only 6 broad-bands versus 20 broad and medium bands. Top row: Parameters obtained from the SED fitting based on 20 medium (MB) and broad bands (BB) as a function of the same parameters measured with the same SFH with only 6 BB filters from the UNCOVER program (F115W, F150W, F200W, F277W, F356W, F444W). From left to right we show the mass-weighted ages, stellar mass, metallicity, extinction, SFR and exponential decline parameter from the SFH. The red lines show the 1:1 relations. Bottom row: Distributions of the uncertainties obtained on these parameters with the two different set of filters. The errors are shows in log scale for the age, stellar mass and SFR. The black histograms show the error measured with only 6 BB filters and the grey ones the uncertainties obtained with 20 BB+MB filters. The grey and black vertical lines show the median value of each distribution. }

    \label{fig:compare_BB_MB}
\end{figure*}

\begin{figure*}
	\includegraphics[width=18cm]{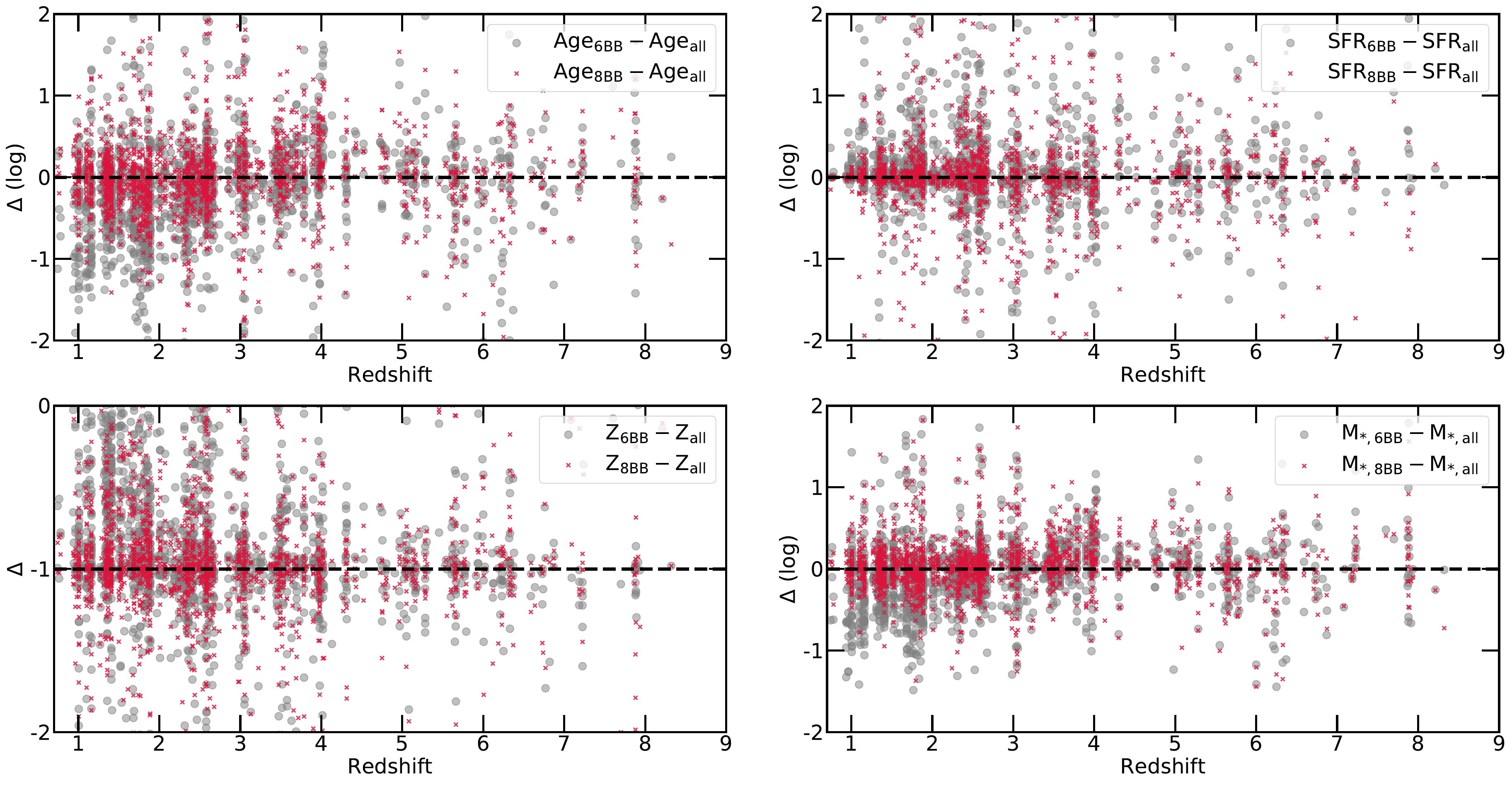}
    \caption{Difference between the best-fit values obtained from SED fitting with 6 BB filters (F115W, F150W, F200W, F277W, F356W, F444W) or the 8 BB and all the BB and MB filters (using the same SFH) in grey and red, respectively, as a function of redshift for the clumps mass-weighted age, stellar mass, SFR and metallicity. The difference is showed in log scale for the age, stellar mass and SFR. The black dashed line show when the two values are equal.}

    \label{fig:compare_BB_MB_redshift}
\end{figure*}

In this section we present the main clumps properties obtained from Bagpipes SED fitting based on the full set of 20 NIRCam filters available (see Table~\ref{tab:table_observations}) versus the same fit based only on the 6 broad-band filters (F115W, F150W, F200W, F277W, F356W, F444W) observed by the UNCOVER program  (GO 2561, P.Is. Labbé and Bezanson, \citealt{Bezanson2022}) . We performed the fit with the reference SFH for the clumps used in our study (i.e., an exponential declining SFH with $\tau=10$ Myr, see Sect.~\ref{sec:clumps_SED}). Fig.~\ref{fig:compare_BB_MB} presents the comparison between the two fit results and uncertainties for the age, stellar mass, metallicity, extinction and SFR. The scattering of the distributions that the best-fit parameters are strongly affected by the addition of two BB and 12 MB filters. Thanks to the addition of 14 filters, the best-fit parameters uncertainties decrease from 0.07, 0.05 and 0.11 (in log scale) for clumps age, mass and SFR, respectively and from 0.009 and 0.04 in metallicity and extinction, respectively. As shown in the Fig.~\ref{fig:fig_filters}, the combination of MB and BB filters allows to constrain both the continuum and individual emission lines EW which will provide more reliable SED best-fit parameters and reduce the uncertainties. The Fig.~\ref{fig:compare_BB_MB_redshift} shows the difference between the clumps best-fit parameters measured with all BB and MB filters and only 6 or 8 BB filters as function of redshift. We observe a larger scatter when using only 6 BB filters than 8, especially at $z<4$ for the age ad SFR. This shows that adding observations with the two filters F070W and F090W are very important to estimate ages and SFR at $z<4$. On average, at all redshift, best-fit SED parameters are significantly affected by the addition of F070W and F090W and MB filters observations. 50\% of the stellar masses are changed from more 0.2 dex in log scale by the addition without F070W, F090W and MB and 30\% without MB. 

\bsp	
\label{lastpage}
\end{document}